\documentclass{article}

\usepackage{tabularx, array, booktabs, calc, fullpage, color, graphicx, epstopdf, floatrow, multirow, lscape, longtable, alphalph, enumitem, amssymb, pifont, tabu}
\usepackage{amsmath}
\usepackage{amsthm}
\usepackage{amsfonts}
\usepackage{graphicx}
\usepackage{listings}
\usepackage{caption}
\usepackage{subcaption}
\usepackage{url}
\usepackage{booktabs}

\usepackage[T1]{fontenc}
\usepackage[utf8]{inputenc}
\usepackage[font=small,labelfont=bf]{caption}

\definecolor{mygreen}{rgb}{0,0.6,0}
\definecolor{mygray}{rgb}{0.5,0.5,0.5}
\definecolor{mymauve}{rgb}{0.58,0,0.82}

\begin{document}

\title{An Analysis of the Twitter Discussion \\ on the 2016 Austrian Presidential Elections}
\author{Ema Ku\v{s}en$^1$ and Mark Strembeck$^{1,2,3}$ \\$^{1}$ \textit{Vienna University of Economics and Business (WU Vienna)} \\$^2$ \textit{Secure Business Austria Research Center (SBA)}\\$^3$ \textit{Complexity Science Hub Vienna (CSH)}}
\date{}
\maketitle

\begin{abstract}
In this paper, we provide a systematic analysis of the Twitter discussion on the 2016 Austrian presidential elections. In particular, we extracted and analyzed a data-set consisting of 343645 Twitter messages related to the 2016 Austrian presidential elections. Our analysis combines methods from network science, sentiment analysis, as well as bot detection. Among other things, we found that: a) the winner of the election (Alexander Van der Bellen) was considerably more popular and influential on Twitter than his opponent, b) the Twitter followers of Van der Bellen substantially participated in the spread of misinformation about him, c) there was a clear polarization in terms of the sentiments spread by Twitter followers of the two presidential candidates, d) the in-degree and out-degree distributions of the underlying communication network are heavy-tailed, and e) compared to other recent events, such as the 2016 Brexit referendum or the 2016 US presidential elections, only a very small number of bots participated in the Twitter discussion on the 2016 Austrian presidential election.
\end{abstract}

\section{Introduction}

Social media platforms such as Facebook, YouTube, or Twitter provide
public communication channels that enable every individual, company,
political party, or government agency to directly get in touch with
each other via text messages, pictures, or videos.  Thereby, social
media break the communication patterns of more traditional media (such
as TV channels or newspaper publishers) by democratizing targeted mass
communication to a certain degree. However, aside from sharing
messages with friends, family, fans, or customers, social media are, at an increasing rate, also used in political campaigns
\cite{twitter-use-by-us-congress,
new-platform-old-habits-twitter-2010-british-and-dutch-election,
rise-of-twitter-in-political-campaign} and for the decentralized
coordination of political protests
\cite{social-media-evolution-egyptian-revolution,
analyzing-impact-of-social-media-twitter-occupy-wallstreet,
osns-and-offline-protest}, for example.

Moreover, because rumors are spreading quickly in social networks
\cite{why-rumors-spread-so-quickly-in-social-networks}, social media
are also a tool for spreading misinformation such as political
propaganda \cite{cacm-info-and-misinfo-on-the-internet}. While it is
not exactly a new finding that information found on the internet is
not reliable per se \cite{cacm-of-course-its-true-i-saw-it-on-the-internet}, the matter becomes more and more important due to the heavy use of online
(mis)information campaigns and phenomena such as social bots (autonomous
software agents for social media platforms) \cite{cacm-rise-of-social-bots,
social-bots-distort-2016-us-election}. In addition to social bots that might try to influence human users, one must also take into account how community feedback shapes user behavior \cite{how-community-feedback-shapes-user-behavior} and consider the significant role that official social media accounts play in an online debate \cite{keeping-up-with-the-tweet-dashians}. Furthermore, and aside from the properties of a particular message source, a thorough analysis has to take the emotional dimension of a discussion into account \cite{emotional-valence-shifts-asonam-2017, word-emotion-lexicons-snams-2017}. For example, a heated debate over a controversial topic develops more dynamically and unpredictably than an objective discussion and might therefore be more easily manipulated. In this context, sentiment analysis methods \cite{Liu2016} help classify and understand the users' emotions.

It was already in 2013 that the World Economic Forum called misinformation in social networks "digital wildfires" which cause a global risk and may lead to serious societal, economic, and political consequences by endangering democracy or influencing markets \cite{WEF2013}. Thus, in a world involving a wide variety of reliable as well as unreliable message sources, "alternative facts", and fake news \cite{post-truth-guide-for-the-perplexed}, scientists of all
disciplines have an obligation to provide the public with the tools and the information to separate fact from fiction
\cite{give-public-tools-to-trust-scientists, take-the-time-and-effort-to-correct-misinformation}. However, the sheer complexity of socio-technical systems \cite{complex-adaptive-systems-book, complexity-a-guided-tour-book} and the big data characteristics of complex networks \cite{big-data-in-complex-and-social-networks-book} make the analysis of social media events a difficult task. Therefore, a number of research projects, such as the DARPA Twitter bot challenge \cite{ieee-computer-darpa-twitter-bot-challenge}, aim to improve our
capabilities for the analysis and understanding of social media
campaigns \cite{power-of-social-media-analytics,
online-deception-in-social-media}. In this context, case studies of
real-world political campaigns are of particular interest because they help
understand human behavior, detect patterns, and identify generic
approaches for analyzing user behavior in online social networks (see,
e.g., \cite{new-platform-old-habits-twitter-2010-british-and-dutch-election, studying-political-microblogging-2010-swedish-election,
Song2014,
alchemy-of-authenticity-lessons-from-2016-us-campaign,
2014-indian-elections-on-twitter-strategies,
explaining-donald-trump-via-communication-style}).

In the 2016 Austrian presidential elections, Austria has witnessed two polarizing opinions among its citizens. A candidate of the Freedom Party of Austria, Norbert Hofer, and his opposing candidate, a former member of the Green Party, Alexander Van der Bellen were in a tight run for the presidential seat. The first round of the elections took place on April 24th 2016, when Norbert Hofer received a majority of the votes (36.40\%), followed by Alexander Van der Bellen (20.38\%), while four other candidates (Irmgard Griss, Rudolf Hundstorfer, Andreas Khol, and Richard Lugner) dropped out of the elections. The second round, which took place on May 22nd 2016, was a run-off ballot between Hofer and Van der Bellen. Alexander Van der Bellen won with 50.3\% of the votes \cite{BMI2016}. However, the results of this election have been invalidated by the Austrian constitutional court in July 2016 due to procedural irregularities in vote counting\footnote{Note that on July 1st 2016, Austria's constitutional court ruled that the presidential election must be repeated due to irregualrities and formal errors in the counting procedures for postal votes in 14 voting districts. As a result of those errors, there was an abstract chance of voter fraud. Evidence of actual voter fraud has not been found though.}. After the re-elections were postponed due to faulty glue on the envelopes for postal voting \cite{Connolly2016}, the repeat of the run-off ballot finally took place on December 4th 2016, when Van der Bellen was elected president with 53.8\% of the votes \cite{BMI22016}. The inauguration ceremony took place on January 26th 2017.

In this paper, we provide a comprehensive analysis of the Twitter discussion related to the 2016 Austrian presidential elections. In particular, we extracted and analyzed a data-set consisting of 343645 Twitter messages 
The resulting data-set is multi-dimensional, including temporal data, structural data (such as the underlying communication network or the corresponding topic/hashtag network), as well information on the user's emotions that are expressed in the content of the messages. In order to provide an in-depth analysis, we therefore combined methods from network science \cite{Newman2010, social-and-economic-networks-book}, sentiment analysis \cite{Liu2016}, as well as bot detection \cite{cacm-rise-of-social-bots, ieee-computer-darpa-twitter-bot-challenge}.

The remainder of this paper is structured as follows. In Section \ref{sec:method}, we provide an approach synopsis and discuss the guiding research questions for our study. Subsequently, Section \ref{sec:data_analysis} presents a systematic analysis of the Twitter discussion on the 2016 Austrian presidential elections. In Section \ref{sec:discussion}, we further discuss our findings, the general approach for analyzing social media events, as well as the limitations of our study. Section \ref{sec:RelatedWork} discusses related work, before Section \ref{sec:conclusion} concludes the paper.

\section{Research Questions and Approach Synopsis} \label{sec:method}

\subsection{Research questions} \label{sec:research_questions}

We defined the following guiding research questions for our analysis:

\medskip
\textit{RQ1: What is the tweeting behavior of the presidential candidates?}

In specific, we examined three aspects: temporal characteristics of each candidate's tweeting behavior (RQ1.1), each candidate's engagement style (RQ1.2), as well as each candidate's campaigning style (RQ1.3).

\medskip
\textit{RQ1.1: What are the temporal characteristics of each candidate's tweeting behavior?}

Research question RQ1.1 provides a quantitative analysis of the tweeting behavior and examines how many daily tweets have been posted by a candidate during the presidential elections. For example, we identify associations between important events (such as a TV discussion) and the corresponding tweet count. 

\textit{RQ1.2: What is the engagement style of each candidate ?}

In research question RQ1.2, we focus on the way each candidate uses Twitter as a tool for communication with their supporters. In particular, we investigated each candidate's interaction with their followers, including the ratio between the candidates' broadcasting behavior and bilateral (one-to-one) communication. In addition to the quantitative analysis of the engagement styles, we also examine the content of the candidates' tweets and report on the emotions they spread during their presidential campaign -- the emotions have been identified by using sentiment analysis techniques (see Section \ref{sec:sentiment_candidates}). Furthermore, we examine the reactions of Twitter users on the candidates' tweets in terms of re-tweets, responses, and likes.

\medskip
\textit{RQ1.3: Is there evidence of different types of campaigning ?}

Political campaigns are generally described as "positive" or "negative", depending on how the candidates address their opponents. In our study, we follow the definition from \cite{Pattie2011}, which describes \textit{negative campaigning} as a type of campaigning which may involve misinformation, ``dirty tricks'', attacks on the opponent's persona (also called \textit{political character assassination}), or stress his/her weaknesses or failures from the past. In contrast, \textit{positive campaigning} disseminates information about a candidate's positive future plans or his/her past success. For example, the use of negative campaigning has been well-documented by reputable media during the 2016 US presidential elections (see, e.g., \cite{social-bots-distort-2016-us-election}). Even though this campaigning strategy prospectively contributed to the success of the Republican candidate (Donald Trump), there is evidence that negative campaigning is risky and might backfire, leading to undesired effects (e.g.\ by making a candidate less likeable, see \cite{Pattie2011}). 
As part of our study, we examined cases of negative campaigning found in our data-set (including the spread of misinformation and rumors) and the effects on the candidates' followers. We do this by 1) searching for known false accusations in our data-set and 2) analyzing the opinion polarities a candidate uses to address the opposing candidate (i.e.\ does the candidate mention his rival in a positive or a negative context). 

\medskip
\textit{RQ2: What is the tweeting behavior of users tweeting about 2016 Austrian presidential elections ?}\\

\textit{RQ2.1: In which context do followers mention the candidates ?}\\
Here we refer to the context in which ("ordinary") Twitter users addressed both candidates. In particular, we used network analysis techniques (see, e.g., \cite{Newman2010, social-and-economic-networks-book, graph-based-social-media-analysis-book, complex-social-networks-book}) to derive and analyze ego-networks of hashtags for each candidate. 

\medskip
\textit{RQ2.2: What are the communication patterns among the candidates' followers ?}

Here, we examine whether communities exist among each candidate's followers. In particular, we derive a communication network and analyze whether users tend to communicate with the followers of the opposing candidate or within their group only. 

\medskip
\textit{RQ3: Is there evidence of bots participating in the Twitter discussion ?}

Here we investigate whether bots are participating in the Twitter discourse, and if so, how they relate to different aspects of information sharing over Twitter. In particular, this refers to the reactions of human users on bots, as well as the contents disseminated by bots (see, e.g., \cite{cacm-rise-of-social-bots, ieee-computer-darpa-twitter-bot-challenge, Davis2016}).

\subsection{Approach synopsis} \label{sec:approach}

In order to get a comprehensive picture of the Twitter discussions during the 2016 Austrian presidential elections, we examined the tweeting behavior of the two candidates (Alexander Van der Bellen and Norbert Hofer)\footnote{In particular, we analyzed messages sent from the @vanderbellen and @norbertghofer Twitter accounts. It is not possible, however, to determine if a particular message was sent by one of the candidates or by some member of their respective social media teams.} and analyzed how their tweeting strategy influenced the tweeting behavior of their respective followers as well as the followers of the opposing candidate. In this context, we define \textit{tweeting behavior} as sending a new tweet, replying to a tweet, liking another user's tweet, and re-tweeting an existing message.

\begin{figure}[ht]
	\centering
		\includegraphics[scale=0.8]{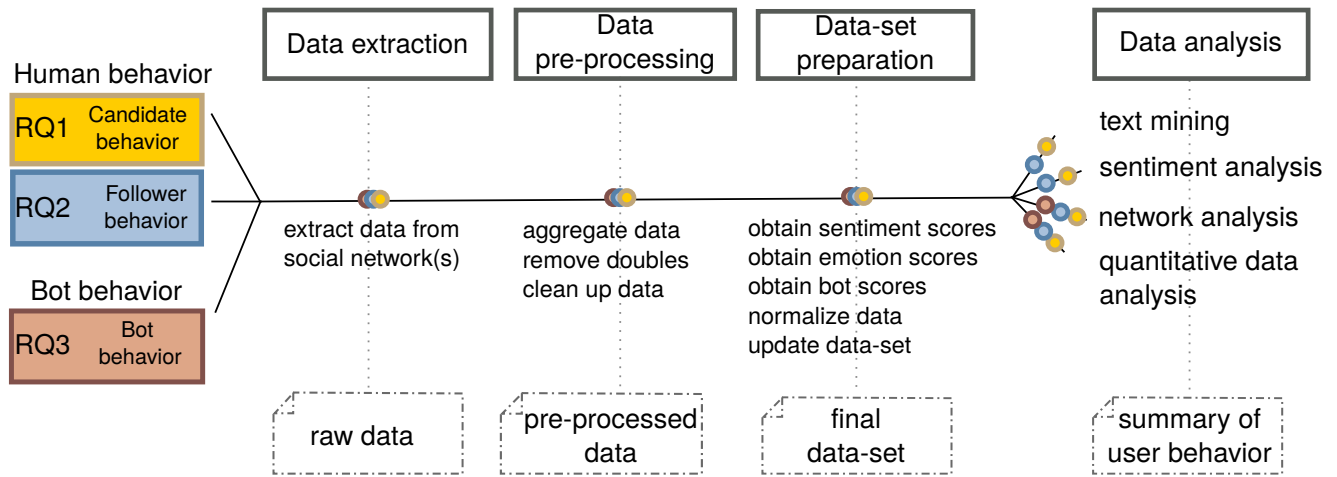}
		\caption{Approach overview: Analyzing social media events}
		\label{fig:method_sketch}
\end{figure}

To study the tweeting behavior, we differentiate between the candidates' tweeting behavior, the tweeting behavior of their followers, as well as the tweeting behavior of autonomous software programs (bots). Our analysis follows three main research questions (see Section \ref{sec:research_questions}) and has been separated in four main phases (see Figure \ref{fig:method_sketch}). 

\textbf{Phase 1 - Data extraction:} In the data extraction phase, we used Twitter's API\footnote{https://dev.twitter.com/overview/api} to collect tweets about the 2016 Austrian Presidential Election. In particular, we collected German language and English language tweets for the run-off election that took place at December 4th 2016. We started the data extraction procedure on November 14th 2016 (three weeks before the election) and continued the extraction procedure until December 14th 2016 (10 days after the election). Even though the official language in Austria is German, we were also interested in English language tweets to capture the opinion of foreigners living in Austria as well as people interested in the elections who were living outside of the country. The data extraction procedure resulted in a data-set consisting of 343766 tweets, 206372 of which are English language tweets and 136372 are German language tweets. Moreover, from March 1st 2016 till December 14th 2016 we also extracted the tweets directly issued by the two presidential candidates, giving us 602 tweets posted by Alexander Van der Bellen (@vanderbellen) and 420 tweets posted by Norbert Hofer (@norbertghofer). The 343766 tweets included 121 double entries (see below), giving us a total of 343645 unique tweets. In order to extract relevant tweets from the Twitter message stream, we thoroughly examined the hashtags used by each campaign and then applied the following list of hashtags for filtering: \textit{\#vdb, \#vdb16, \#VanDerBellen, \#MehrDennJe, \#NorbertHofer, \#NorbertHofer2016, \#Hofer, \#bpw16, \#AustrianElection}, as well as combined occurrences of \textit{\#Austria and \#election}. For each of the 343645 tweets, we extracted the following information: 

\begin{itemize}[noitemsep]
\item the text body of the tweet, 
\item the corresponding Twitter username, 
\item the time and date when the tweet has been published, 
\item the corresponding re-tweet count, 
\item and the corresponding ``like'' count. 
\end{itemize}

For the tweets of the two presidential candidates, we also extracted the reply count for each of the candidate's tweets and a list of their followers (Twitter IDs and usernames).

\textbf{Phase 2 - Data pre-processing:} In the data pre-processing phase, we dealt with aggregating and encoding the raw data gathered in Phase 1. Thus, we inserted new columns into the data-set -- e.g.\ the "follower" column includes discrete values from "0" to "3", where 0 meant that the respective Twitter user does not follow any candidate, 1 encodes followers of Van der Bellen, 2 encodes followers of Norbert Hofer, and 3 encodes followers of both candidates. Moreover, we added the reply count to the candidates' data-sets. Since we used the above list of hashtags to extract our data, tweets that use a combination of those hashtags might have produced double entries in our data-set. For example, the following three hashtags have been combined particularly often  in the same tweet: \textit{\#vdb},\textit{\#bpw16}, and \textit{\#norberthofer}. Therefore, we applied a data cleaning procedure to identify and remove 121 double entries from our data-set. Moreover, since people freely express themselves on Twitter, the language is not formal and contains spelling errors, abbreviations, alternative spelling, and slang words, which, if not addressed properly, might cause errors in a subsequent data analysis (see, e.g., \cite{data-cleaning-detecting-diagnosing-editing-data-abnormalities}). Thus, in order to normalize the extracted data we manually searched for and adjusted typing errors (e.g.\ we corrected ``van der Belen'' to ``Van der Bellen'', ``Östereich'' to ``Österreich'', etc.) or alternative spelling (e.g.\ we replaced ``Oesterreich'' with ``Österreich'', ``Praesident'' with ``Präsident'', etc.).

\textbf{Phase 3 - Data-set preparation:} In this phase, we ran the data-set through SentiStrength \cite{Thelwall2010} and the \textit{BotOrNot} Python API \cite{Davis2016} to obtain sentiment scores for each tweet and bot scores for each username. In addition, we also applied the NRC emotion-word dictionary \cite{Mohammad13} over the tweets and stored emotions identified in the tweets. These scores were then added to our data-set (see also Section \ref{sec:data_analysis}).  

\textbf{Phase 4 - Data analysis:} In the analysis phase, we conducted our data analysis over the final version of the data-set (see also Figure \ref{fig:method_sketch}). In particular, we used text mining techniques, sentiment analysis, network analysis, and quantitative data analysis to find patterns in the user's tweeting behavior (see Section \ref{sec:data_analysis}).

\textbf{Software tools:} For data extraction, pre-processing, and data analysis, we used R\footnote{https://www.r-project.org/}, as well as the follwing R packages: \textit{igraph}\footnote{http://igraph.org/}, \textit{stringr}\footnote{https://cran.r-project.org/web/packages/stringr/}, and \textit{tm}\footnote{https://cran.r-project.org/web/packages/tm/}. Furthermore, we used the SentiStrength\footnote{http://sentistrength.wlv.ac.uk/} tool and the NRC dictionary\footnote{http://saifmohammad.com/WebPages/NRC-Emotion-Lexicon.htm} for extracting sentiment polarities and emotion vectors. Finally, we used the \textit{BotOrNot} Python API\footnote{https://github.com/truthy/botornot-python/} for bot detection purposes.

\section{Data Analysis} \label{sec:data_analysis}

\subsection{Tweeting behavior of the presidential candidates (RQ1)}\label{sec:rq1}

\subsubsection{Temporal properties of the candidates' tweeting behavior}

\begin{minipage}{\linewidth}
      \centering
      \begin{minipage}{0.45\linewidth}
          \begin{figure}[H]
              \includegraphics[scale=0.4]{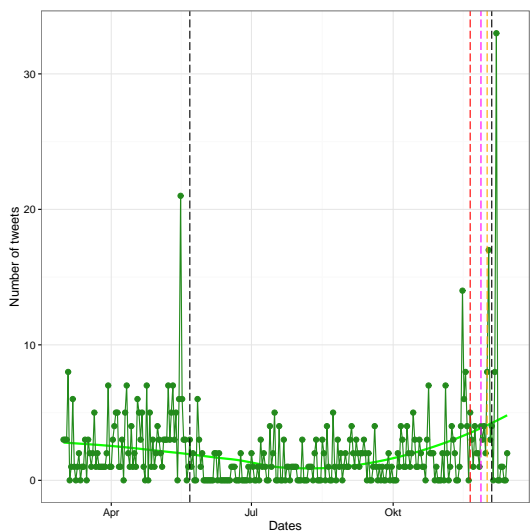}
              \caption{Van der Bellen's tweeting timeline.}
							\label{fig:vdb_tweeting_timelines}
          \end{figure}
      \end{minipage}
      \hspace{0.05\linewidth}
      \begin{minipage}{0.45\linewidth}
          \begin{figure}[H]
              \includegraphics[scale=0.4]{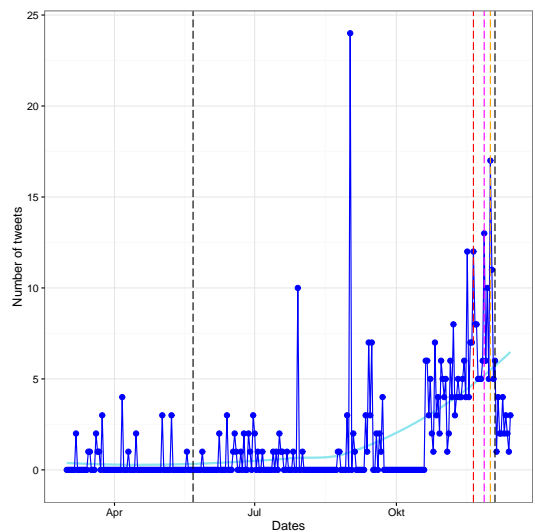}
              \caption{Hofer's tweeting timeline.}
							\label{fig:nh_tweeting_timelines}
          \end{figure}
      \end{minipage}
  \end{minipage}

Our data-set includes tweets that the presidential candidates posted between March 1st 2016 and December 14th 2016 (see Section \ref{sec:approach}). Figures \ref{fig:vdb_tweeting_timelines} and \ref{fig:nh_tweeting_timelines} show the tweet count per day for each candidate, with dashed lines indicating important real-world events that happened during the campaign. In particular, the two black dashed lines mark the election days (May 22nd and December 4th 2016), the orange, magenta, and red lines each mark the dates of different TV discussions between the candidates respectively\footnote{The red line marks the discussion broadcast on PULS4 (November 20th), the magenta line marks the discussion broadcast on ATV (November 27th), and the orange line marks the date of a TV debate broadcast on ORF2 (December 1st)}. The plots in Figures \ref{fig:vdb_tweeting_timelines} and \ref{fig:nh_tweeting_timelines} show the tendency of the candidates to increase their tweeting activity shortly before an important event. Such behavior has also been observed in other elections in Europe (see, e.g., \cite{Smailovic2015}). In our data-set, this trend is particularly evident in Van der Bellen's tweeting timeline. In contrast, Norbert Hofer's tweeting activity was comparatively low during the first round of elections but increased considerably since October 2016. Moreover, Figure \ref{fig:nh_tweeting_timelines} shows a  peak on September 1st. Since no important political event (such as an election day or a TV discussion) took place around that date which would explain such an increase in the tweet count, we manually examined the content of the corresponding tweets. In this particular case, the candidate responded to negative tweets that were directed at him.

\subsubsection{Engagement style of the presidential candidates}

In addition to the tweeting frequency, the message contents also showed that the candidates followed different approaches.  
Tweets originating from Van der Bellen's account frequently used the candidate's first name, the name of the country (\textit{Österreich}, German for ``Austria''), as well as a range of positive words, such as "together" (de: \textit{gemeinsam}), "collaboration" (de: \textit{Zusammenarbeit}), "support" (de: \textit{unterstützen}), as well as informative words about his presence in the media (de: \textit{Gast}, \textit{Interview}, \textit{Plakatpräsentation}). Tweets originating from Norbert Hofer's account used a more personal approach to address his supporters. The tweets often started with the term ``\textit{dear friends}'' (de: \textit{liebe Freunde}) and ended with ``\textit{yours Norbert}'' (de: \textit{Euer Norbert}). 

\textit{Sentiment analysis} \cite{Liu2016} is concerned with studying people's opinions, attitudes, and emotions by analyzing written text. For example, the corresponding techniques are used for analyzing public opinion during elections (see, e.g., \cite{Diaz2012}), for identifying potential radicalization suspects (\cite{Bermingham2009}), or even for predicting stock market events (see, e.g., \cite{Bollen2011,Zheludev2014}).

In order to gain more insight into the emotions and sentiment polarities both candidates target in their followers, we analyzed the sentiment polarities by using the SentiStrength algorithm \cite{Thelwall2010}. Moreover, we identified 8 basic emotions according to Plutchik's wheel of emotions \cite{Plutchik2001} by applying the NRC lexicon \cite{Mohammad13} over the candidates' tweets. Our analysis was based on the assumption that a single tweet may contain positive emotions, negative emotions, or a mixture of both. 

Figures \ref{fig:vdb_senti} and \ref{fig:nh_senti} summarize the sentiment polarities we identified in each candidate's tweets. In particular, we grouped the tweets into one of four categories respectively: positive, negative, neutral, or  overlap between positive and negative polarity. This classification was done by using the sentiment scores given by the SentiStrength algorithm. 

Figures \ref{fig:vdb_senti} and \ref{fig:nh_senti} show a substantial difference in sentiment polarities as communicated by each candidate. Van der Bellen mostly posted neutral tweets where he announced TV debates, radio talk shows, and other pre-election events. In addition, a number of positive tweets originated from Van der Bellen's account, with their number increasing on election day when the candidate expressed his gratitude and thanks to his supporters. While the number is comparatively low, Van der Bellen's tweets also include some tweets with negative sentiment polarity scores. Those negative messages mostly refer to negative events that happened worldwide, such as bombings in Istanbul (Turkey) that took place on December 10th 2016. In comparison, tweets originating from Norbert Hofer's account are more emotionally driven. In particular, he shared comparatively more tweets with positive and negative emotions, such as his love for the country, and gratitude to his supporters (positive), as well as tweets with a negative content (such as answers to negative tweets about himself and his opinion on terrorist attacks).

\begin{minipage}{\linewidth}
      \centering
      \begin{minipage}{0.45\linewidth}
          \begin{figure}[H]
              \includegraphics[scale=0.5]{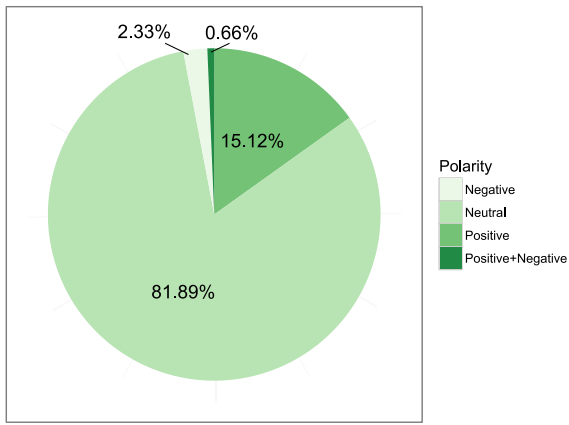}
              \caption{Sentiments in Van der Bellen's tweets.}
							\label{fig:vdb_senti}
          \end{figure}
      \end{minipage}
      \hspace{0.05\linewidth}
      \begin{minipage}{0.45\linewidth}
          \begin{figure}[H]
              \includegraphics[scale=0.5]{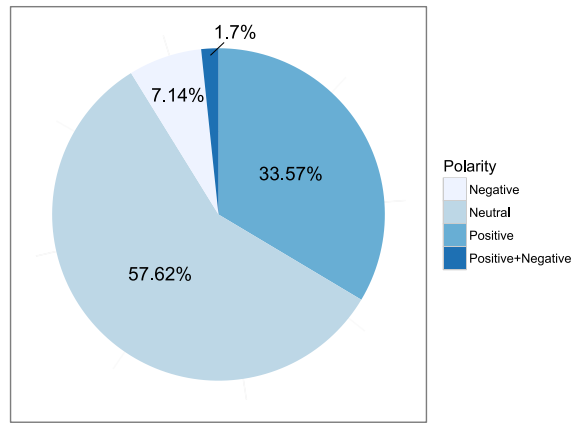}
              \caption{Sentiments in Norbert Hofer's tweets.}
							\label{fig:nh_senti}
          \end{figure}
      \end{minipage}
  \end{minipage}

\bigskip

\begin{figure}[H]
      \includegraphics[scale=0.5]{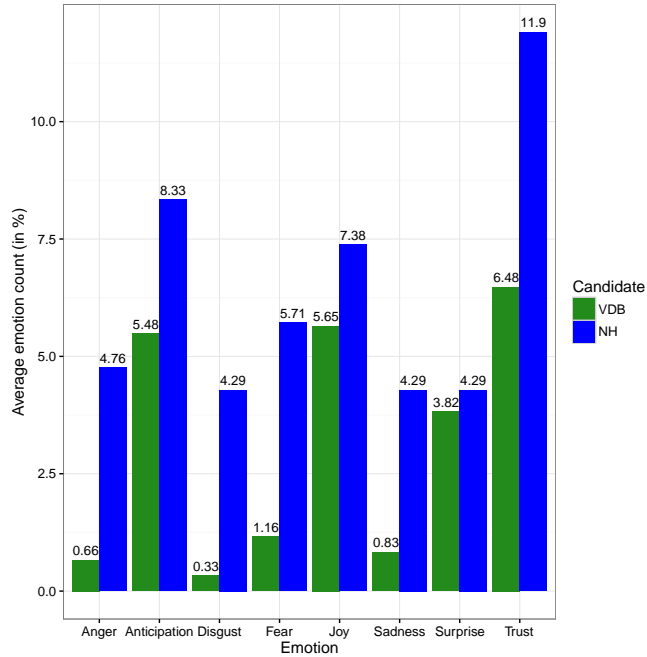}
      \caption{Emotions communicated by both presidential candidates.}
			\label{fig:nh_vdb_senti}
  \end{figure}

To further examine which emotions contribute to the positive and negative sentiment scores, we used the NRC lexicon to identify eight basic emotions according to Plutchik's wheel of emotions. In particular, we found anger, disgust, fear, and sadness for negative SentiStrength sentiment scores as well as joy and trust for positive SentiStrength scores. 
The results shown in Figure \ref{fig:nh_vdb_senti} indicate that even though Van der Bellen posted more tweets in total, his tweets were not as emotionally charged as those posted from Norbert Hofer's account.

The results further indicate a larger difference in the range of negative emotions (anger, sadness, disgust, and fear) than in positive emotions (joy and trust). Therefore, after examining the emotions communicated by both candidates, we can identify a noticeable distinction in the communication strategy between both candidates. While Van der Bellen posted predominantly messages with a neutral sentiment score (esp.\ tweets of informative nature), Norber Hofer addressed his supporters with more emotional and personal tweets.
	
To promote their campaign over Twitter in terms of hashtags, the two candidates again took different approaches. 
Tweets that originated from Van der Bellen's account used the hashtag \textit{\#vanderbellen} in their tweets, while tweets originating from Norbert Hofer's account did not use his own name as a hashtag. Instead, Norbert Hofer used hashtags to indicate important TV discussions (\textit{\#dasduell, \#orfduell, \#puls4, \#atv}). Furthermore, both candidates used the official Austrian presidential election hashtag \textit{\#bpw16} (an acronym for German: ``Bundespr\"asidentenwahl 2016'').
		
\begin{figure}[H]
    \includegraphics[scale=0.4]{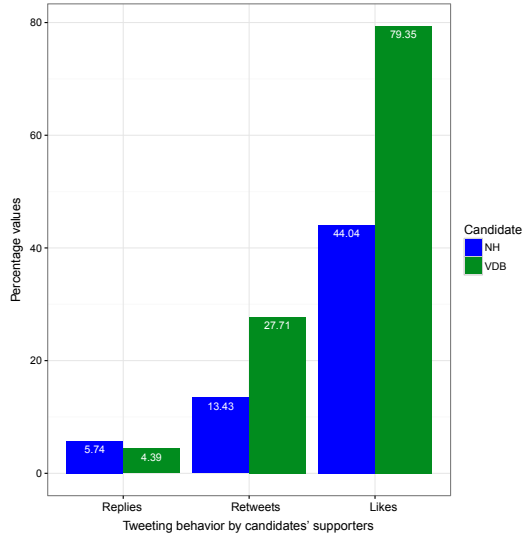}
    \caption{Summary of the tweeting behavior separated by followers of each candidate.}
		\label{fig:vdb_nh_summary}
\end{figure}

Next, we analyze the reactions of other ("ordinary") Twitter users to the tweets posted by both candidates. In particular, we examined the number of replies, re-tweets, and likes averaged over the total number of tweets posted by each candidate (see Figure \ref{fig:vdb_nh_summary}). The results indicate that, in general, Norbert Hofer inspired more replies to his tweets, while Van der Bellen's tweets received more likes and re-tweets. 

Because two Twitter users can directly communicate with each other by using the @ character followed by the other user's username, we were able to trace such direct communication. In total, 35.05\% of the tweets posted by Van der Bellen are messages directed to another user. These tweets predominantly occurred towards the end of the data extraction period when the candidate thanked users for their congratulation messages after winning the presidential elections. In contrast, only 15\% of the tweets originating from Norbert Hofer's account are direct responses to another user's tweets.

\begin{figure}[ht]
	\centering
		\includegraphics[scale=0.7]{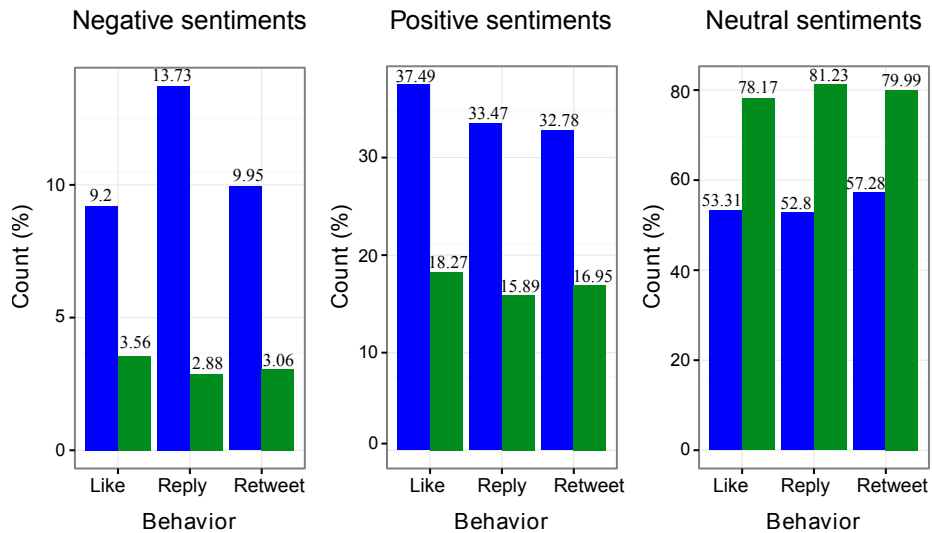}
		\caption{Tweeting behavior for positive and negative sentiment scores (Hofer: blue, Van der Bellen: green)}
	\label{fig:sentiment_behavior}
\end{figure}

In our analysis, we also observed a strong positive correlation between the re-tweet-count and the like-count for both presidential candidates. Pearson's coefficient for re-tweets of Norbert Hofer's messages and corresponding likes is a strong positive 0.94, while Van der Bellen's messages show a correlation coefficient of 0.95 between re-tweets and likes. 
Figure \ref{fig:sentiment_behavior} shows that, compared to tweets from Van der Bellen's account, tweets originating from Norbert Hofer's account generally received considerably more re-tweets, replies, and likes if the respective message expressed positive or negative emotions.

Since the three most replied tweets from both candidates expressed emotions, we also examined if the reply count correlates with emotionally charged tweets. We found that for messages originating from Van der Bellen's account, the association between positive emotions and the reply count is weak and negative (Pearson coefficient equals -0.0012), while tweets with negative emotions exhibit a positive, but weak correlation with the reply count (Pearson's coefficient is 0.023). With respect to messages originating from Norbert Hofer's account, positive emotions and negative emotions both positively correlate with the reply-count (Pearson's coefficient for positive emotions and reply-count is 0.14, while for negative emotions and reply-count Pearson's coefficient is 0.12).

\subsubsection{Analysis of campaign styles with respect sentiment polarities}
\label{sec:sentiment_candidates}

During the 2016 Austrian presidential election, there was some evidence of negative campaigning in the Austrian media (e.g.\ TV discussions where Van der Bellen was accused of being a spy \cite{SpyPresse2016, Spy2016}). In our dataset, we found a subset of tweets posted by each candidate that mention the opposing candidate or his (former) party (FP\"O for Norbet Hofer and Green party for Van der Bellen). In total, 1.82\% (11) of the tweets originating from Van der Bellen's account mention Norbert Hofer, seven of which are neutral (esp.\ announcements for TV or radio discussions with Norbert Hofer), while the rest share a negative sentiment about the opposing candidate. 
In comparison, tweets originating from Norbert Hofer's account mentioned his opponent slightly more often. In particular, 3.57\% (15) of the tweets mention Van der Bellen, five of which are neutral, while the rest directly express some opinion that results in a negative SentiStrength sentiment score about the Green party or one of Van der Bellen's messages. Figures \ref{fig:vdb_mention_hofer} and \ref{fig:nh_mention_vdb} show the emotions found in tweets that each candidate used to address his opponent.
	
	\begin{minipage}{\linewidth}
      \centering
      \begin{minipage}{0.45\linewidth}
         \begin{figure}[H]
       \includegraphics[scale=0.4]{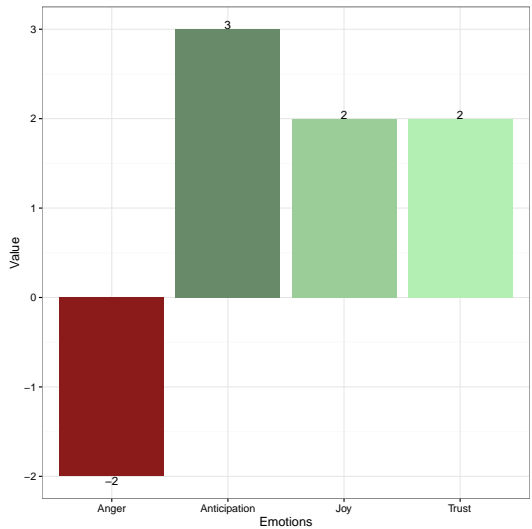}
      \caption{Emotions in Van der Bellen's tweets that mention Hofer}
			\label{fig:vdb_mention_hofer}
			\end{figure}
      \end{minipage}
      \hspace{0.05\linewidth}
      \begin{minipage}{0.45\linewidth}
          \begin{figure}[H]
      \includegraphics[scale=0.4]{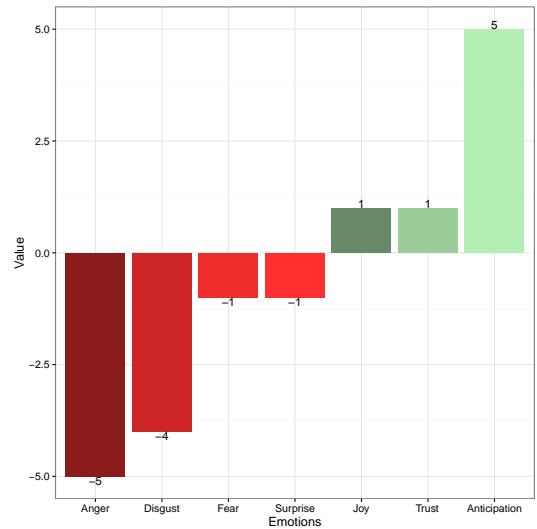}
      \caption{Emotions in Hofer's tweets that mention Van der Bellen}
			\label{fig:nh_mention_vdb}
			\end{figure}
      \end{minipage}
  \end{minipage}

\bigskip

For each candidate, Figures \ref{fig:vdb_about_nh} and \ref{fig:hofer_about_vdb} show the impact of tweets about the opposing candidate in comparison to tweets on other topics. In particular, the plots show the arithmetic mean of the re-tweet count, reply-count, and like-count for messages mentioning the opposing candidate in contrast to the respective numbers for tweets on other topics. In general, tweets in which Van der Bellen mentioned Hofer received slightly more re-tweets and replies than his tweets on other topics. In the same way, Norbert Hofer's tweets that mention Van der Bellen received more replies and re-tweets than his tweets on other topics, however, considerably less likes, compared to other tweets.  

\begin{minipage}{\linewidth}
      \centering
      \begin{minipage}{0.45\linewidth}
         \begin{figure}[H]
       \includegraphics[scale=0.4]{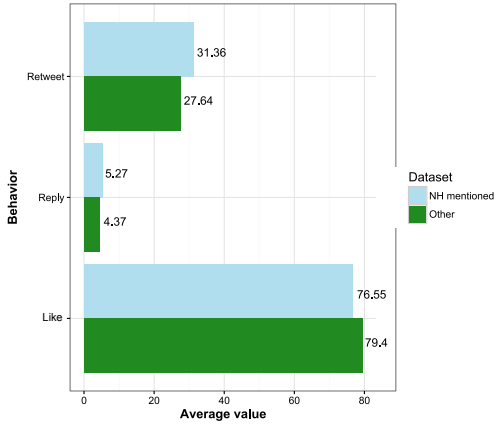}
      \caption{Comparison of the tweets in which Norbert Hofer was mentioned and other tweets posted by Van der Bellen.}
			\label{fig:vdb_about_nh}
			\end{figure}
      \end{minipage}
      \hspace{0.05\linewidth}
      \begin{minipage}{0.45\linewidth}
       \begin{figure}[H]
      \includegraphics[scale=0.4]{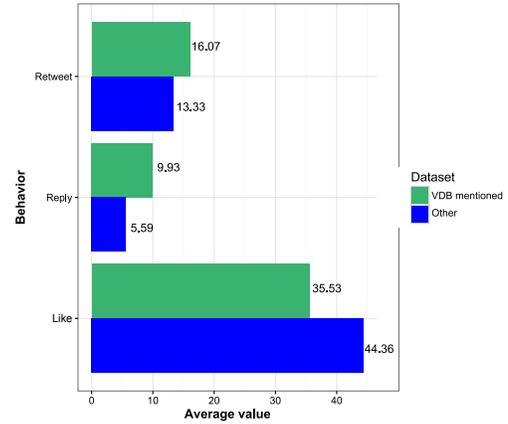}
      \caption{Comparison of the tweets in which Van der Bellen was mentioned and other tweets posted by Norbert Hofer.}
			\label{fig:hofer_about_vdb}
			\end{figure}
      \end{minipage}
  \end{minipage}

\bigskip

\begin{figure}[H]
     \includegraphics[scale=0.4]{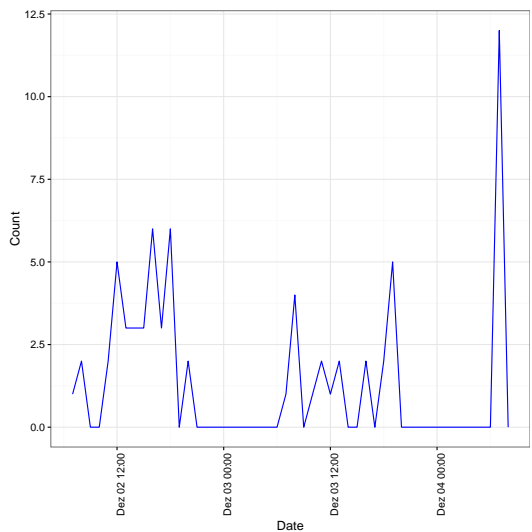}
     \caption{Re-tweet count for the video statement \textit{``Hate in the network and why I am against Van der Bellen.''}}
		\label{fig:hass_neginfo}
	\end{figure}

\textbf{Spread of negative information: }
Figures \ref{fig:fpo_neginfo}, \ref{fig:gertrude_neginfo}, and \ref{fig:hass_neginfo} show three examples of negative campaigning that referred to video statements and got the most re-tweets. In particular, Figure \ref{fig:fpo_neginfo} shows the re-tweet count for the video statement \textit{``Up to now, I always voted for the FPÖ. Why I vote now for \#VanderBellen.''.} The tweet was published by Van der Bellen's official Twitter account on December 1st 2016 at 7:37 AM. Subsequently, it has been re-tweeted and copied over Twitter by 167 distinct followers of Van der Bellen and 2 followers of Norbert Hofer. 

\begin{minipage}{\linewidth}
      \centering
      \begin{minipage}{0.45\linewidth}
         \begin{figure}[H]
       \includegraphics[scale=0.4]{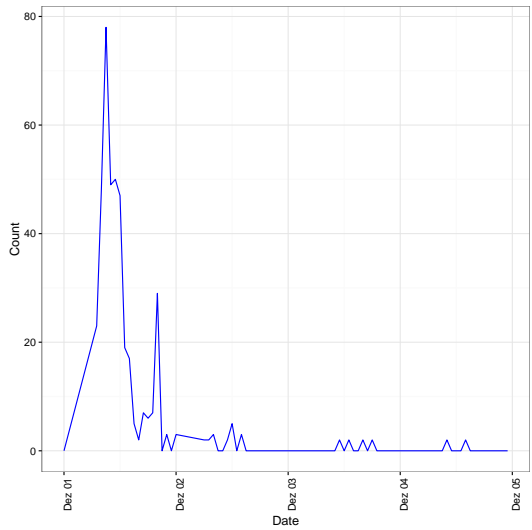}
      \caption{re-tweet count for the video statement \textit{``Up to now, I always voted for the FPÖ. Why I vote now for \#VanderBellen.''.}}
			\label{fig:fpo_neginfo}
			\end{figure}
      \end{minipage}
      \hspace{0.05\linewidth}
      \begin{minipage}{0.45\linewidth}
       \begin{figure}[H]
      \includegraphics[scale=0.4]{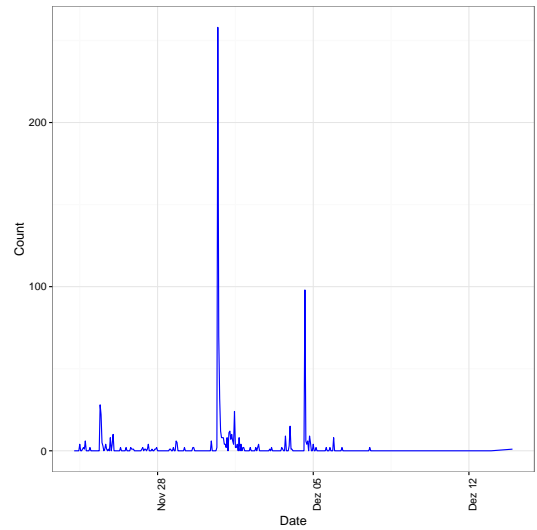}
      \caption{re-tweet count for the video statement \textit{``Gertrude, an 89 year old holocaust survivor warns against FPÖ.''}}
			\label{fig:gertrude_neginfo}
			\end{figure}
      \end{minipage}
  \end{minipage}


A second example is shown in Figure \ref{fig:gertrude_neginfo}, which is another video statement referred to in a tweet that originated from Van der Bellen's account. This video shows an 89-year old holocaust survivor who warns against voting for FPÖ (respectively Norbert Hofer). The tweet was published by the official Twitter account of Van der Bellen on November 24th 2016 at 10:33 AM. Subsequently, it was also disseminated into the English language Twitter-sphere and was re-tweeted and copied 864 times by 135 distinct followers of Van der Bellen and 2 followers of Norbert Hofer. This particular tweet is also the one that continued to spread for the longest time of all tweets that carried negative information about the opposing candidate. Moreover, it is the third most re-tweeted message in our data-set, only preceded by two tweets in which Van der Bellen thanks his supporters for their votes and a tweet which invites people to vote. 

Figure \ref{fig:hass_neginfo} shows the re-tweet count for the video statement \textit{``Hate in the network and why I am against Van der Bellen.''} which was published by one of Norbert Hofer's followers on August 7th 2016 on YouTube. Subsequently, it reached Twitter and was re-tweeted 68 times by 2 followers of Norbert Hofer, and 16 users that do not follow either of the candidates on Twitter. In this example, we witness a higher average re-tweet count (3.78) by each user, whereby one of Norbert Hofer's followers alone was responsible for 12 out of 68 (17.65\%) re-tweets.

\textbf{Spread of misinformation: }In addition to the spread of negative information, the 2016 Austrian presidential elections also witnessed a number of messages including misinformation that spread over Twitter, most of which targeted at Van der Bellen. In particular, the tweets that carry misinformation refer to false accusations of Van der Bellen being a communist spy, and that he suffers from cancer and dementia. We classified this information as misinformation because it has been rebuked by reputable and credible sources, such as quality newspapers including ``Der Standard'' and ``Die Presse'' (see, e.g., \cite{Dementia2016, Krebs2016, Spion2016}). 
Note, however, that neither of the aforementioned examples was posted from Norbert Hofer's Twitter account. Nevertheless, it was mentioned during the candidates' TV discussions and later on discussed and spread over Twitter by (predominantly and surprisingly) Van der Bellen's followers. Thus, Van der Bellen's followers substantially participated in the dissemination of misinformation concerning Van der Bellen. 

In particular, we identified four cases of misinformation in our data-set. In Figure \ref{fig:misinfo}, the red line represents the misinformation about Van der Bellen being a spy. In the plot we can observe that the highest peak of this information stream was reached on December 1st 2016, the date of the ORF2 TV-discussion, when Hofer suggested on TV that Van der Bellen is a spy (see \cite{SpyPresse2016}). In order to gain further insight into people's reactions and behavior over Twitter once they have been exposed to the misinformation, we manually examined the tweets referring to this false ``Spy'' accusation\footnote{Note that in this paper we provide a detailed discussion of one particular misinformation stream - which is the one related to the spy accusations. We chose to present this case in detail because it was the most abundant stream regarding its scope (i.e.\ highest number of reactions and the longest time period).}. 

On the one hand, the corresponding tweets exhibited signs of information seeking (e.g.\ ``\textit{Was Alexander \#VanderBellen a spy?}'' followed by a link to an information source). We found 53 such tweets (including corresponding re-tweets) that were posted a day after the TV discussion and continued spreading for  three days after the discussion. Thus, the misinformation was at first regarded a rumor by some Twitter users that was yet to be confirmed or rebuked. Other users showed signs of annoyance (e.g.\ \textit{``What next is \#VdB going to be?''}) (13 tweets), assumed a threat to Van der Bellen's election success (e.g.\ \textit{``Spy accusations might cost \#VanDerBellen the elections''}) (23 tweets), sarcasm (see, e.g., Figure \ref{fig:vdb_spion} and ``\textit{\#VanDerBond: A spy who loved me.}'' or \textit{``\#VanDerBellen aka spy agent, I always wanted to have James Bond as a president!''}) (36 tweets), and tweets that found the accusation amusing (e.g.\ \textit{``Get your popcorn and turn on @ORF''}) (10 tweets). Although the defending tweets and information seeking tweets were predominant in our data-set, there was also a comparatively smaller number of tweets that confirmed the accusation by providing alleged evidence (e.g.\ \textit{``A book of an ex-security manager is a surprisingly good source: VdB actually was a communist spy.''}), calling the candidate ``ineligible'' (German: \textit{unwählbar}), or criticized Hofer for bringing up the false accusation in the first place (e.g.\ \textit{``Mr. \#Hofer as a president should speak the truth.''}, or via a hashtag \textit{\#liar} next to \textit{\#hofer}). Thus, spreading misinformation as a campaigning strategy has also shown its risks as it partially backfired against the spreader (Hofer).

\begin{minipage}{\linewidth}
      \centering
      \begin{minipage}{0.45\linewidth}
         \begin{figure}[H]
       \includegraphics[scale=0.4]{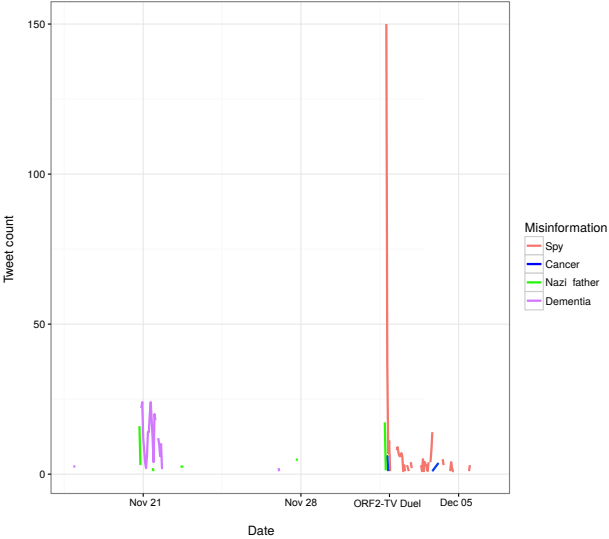}
      \caption{The time-series plot of 4 instances of misinformation spread over Twitter during the presidential elections.}
			\label{fig:misinfo}
			\end{figure}
      \end{minipage}
      \hspace{0.05\linewidth}
      \begin{minipage}{0.45\linewidth}
       \begin{figure}[H]
      \includegraphics[scale=0.2]{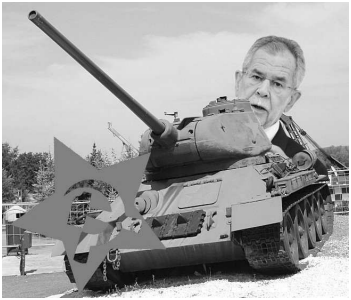}
      \caption{A sarcastic reaction to the tweet suggesting that Van der Bellen is a spy that writes: ``\textit{I just found a top secret photo of Van der Bellen's spying activities}''. This tweet was posted by one of Van der Bellen's followers. It was re-tweeted by two other of his followers as well as two Twitter users that did not follow either candidate on Twitter.}
			\label{fig:vdb_spion}
			\end{figure}
      \end{minipage}
  \end{minipage}


\subsection{Tweeting behavior of other Twitter users (RQ2)} \label{sec:rq2}

In this section, we examine communication patterns and the connectivity among the followers of both presidential candidates. Thus, the subsequent discussion refers to tweets posted by Twitter users other than Van der Bellen and Hofer. We also compare the results that can be obtained from the 136372 German language tweets (subsequently referred to as the "German language data-set") to the corresponding results of the 206372 English language tweets (subsequently referred to as the "English language data-set"). 

\subsubsection{Context of the tweets mentioning each candidate}

First, we were able to confirm that a considerable amount of Twitter content consists of re-tweets. In particular, only 43.1\% of the 136372 tweets in the German language data-set are original tweets, while the remaining 56.9\% are re-tweets. This is even more obvious in the English language data-set, where only 29.89\% of the 206372 tweets are original tweets while 70.11\% are re-tweets. However, this result was expected to a certain degree, since the English Twitter-sphere was predominantly used to disseminate important facts about the Austrian elections, with little to no one-to-one discussion (see also the corresponding network analysis below). In comparison, the German Twitter-sphere witnessed a more extensive discussion about the candidates and the events that happened during the election period. 
 
In order to identify relations between Twitter topics, we derived the corresponding hashtag network. The hashtag network derived from the German language data-set is an undirected network and consists of 5233 distinct vertices and 23535 edges, with an average vertex degree of 9.01. In total, the network includes nine connected components. In particular, some hashtags are isolated and not used in a combination with the hashtags in the giant component of the network. 

The vertices (hashtags) with the largest degree ($\delta$) are \#bpw16 ($\delta$=3872), \#Hofer ($\delta$=1597), \#vdb ($\delta$=1464), \#vanderbellen ($\delta$=1054), and \#Österreich ($\delta$=708). Moreover, it is worth mentioning that in the German language hashtag network, the hashtags \#MarineLePen ($\delta$=325), \#ViktorOrban ($\delta$=293), and \#Trump ($\delta$=233) are among the top fifteen vertices with respect to the vertex degree. The German language hashtag network also shows that both candidates were addressed in a positive and a negative context respectively. 

To examine the context of the discussion related to different hashtags, we also derived two ego-networks, one for each candidate. Figure \ref{fig:de_egonet_hofer} shows Norbert Hofer's (\#hofer) German language ego-network which consists of 1596 vertices and 10846 edges (network density $\approx 0.01$). The German language ego-network of Van der Bellen consists of 1463 vertices and 9878 edges (network density $\approx  0.01$).  In particular, an ego-network consisting of hashtags may reveal valuable insights about the topics people associate with each candidate. After a thorough examination of the hashtags directly connected to each candidate, we identified five categories of hashtags in both ego-networks. Those categories are:

\begin{itemize}[noitemsep]
\item hashtags that directly support a candidate (here we excluded the general hashtag which carries the candidate's name only, because it can either appear in a tweet with negative or positive sentiment polarities), 
\item hashtags that carry general information about the 2016 Austrian presidential elections (e.g. newspaper titles, TV station names, party names, important dates), 
\item hashtags that directly oppose (speak against) a candidate, 
\item hashtags that refer to important topics discussed during the presidential election, 
\item as well as other (hashtags that neither support, go against, carry general information about the elections, or refer to important topics).
\end{itemize}

Below, we list examples from each category:

\begin{enumerate}[noitemsep]
	\item \textbf{Supporting:} e.g. \textit{\#vote4vdb, \#teamvanderbellen}, etc., and \textit{\#Hofer4President, \#hofer2016}, etc.
	\item \textbf{General:} e.g. \textit{\#bpw16, \#presidentialElection, \#norberthofer, \#VanDerBellen}, etc.
	\item \textbf{Against:} e.g. \textit{\#notoVDB, \#VollDerBluff}, and \textit{\#womenAgainsthofer, \#nohofer}, etc.
	\item \textbf{Important topics:} e.g. \textit{\#Islam, \#HillaryClinton, \#Trump, \#terror, \#Brexit, \#Burka}, etc.
	\item \textbf{Other:} e.g. \textit{\#Styria, \#Monday, \#Christmas}, etc.
\end{enumerate}

Figures \ref{fig:de_egonet_vdb} and \ref{fig:de_egonet_hofer} show an extract of the German language ego-networks including the vertices which belong to the categories \textit{supporting, general, against}, and \textit{important topics} (i.e.\ vertices belonging to the "other" category have been excluded from these plots). The corresponding ego-network for Hofer includes 622 vertices and 4826 edges (network density $\approx  0.025$). The respective ego-network of Van der Bellen includes 482 vertices and 3938 edges (network density $\approx  0.034$)\footnote{Note that in the plots the size of a vertex is based on its degree, and the network topology was visualized based on each vertex's community membership. For visualization purposes, we applied the community detection algorithm available in Gephi, as described in \cite{Blondel2008}.}. Vertices in the \textit{Supporting} category are plotted in green color, vertices from the \textit{Against} category are plotted in red, vertices on \textit{Important topics} in yellow, and \textit{General information} in gray.
 
Compared to the German hashtag ego-network, the English language ego-network includes a smaller number of vertices, indicating a lower variety of hashtags.
In particular, the corresponding ego-network of Van der Bellen (see Figure \ref{fig:en_ego_vdb}) includes 131 vertices and 1057 edges (network density $\approx 0.124$). The English language ego-network of Hofer (Figure \ref{fig:en_ego_hofer}) includes 293 vertices and 1927 edges (network density $\approx  0.045$). The low number of unique hashtags (vertices) results from the fact that the English data-set predominantly consists of re-tweets (see also Section \ref{sec:rq2}). 

\begin{minipage}{\linewidth}
      \centering
      \begin{minipage}{0.45\linewidth}
        \begin{figure}[H]
       \includegraphics[scale=0.3]{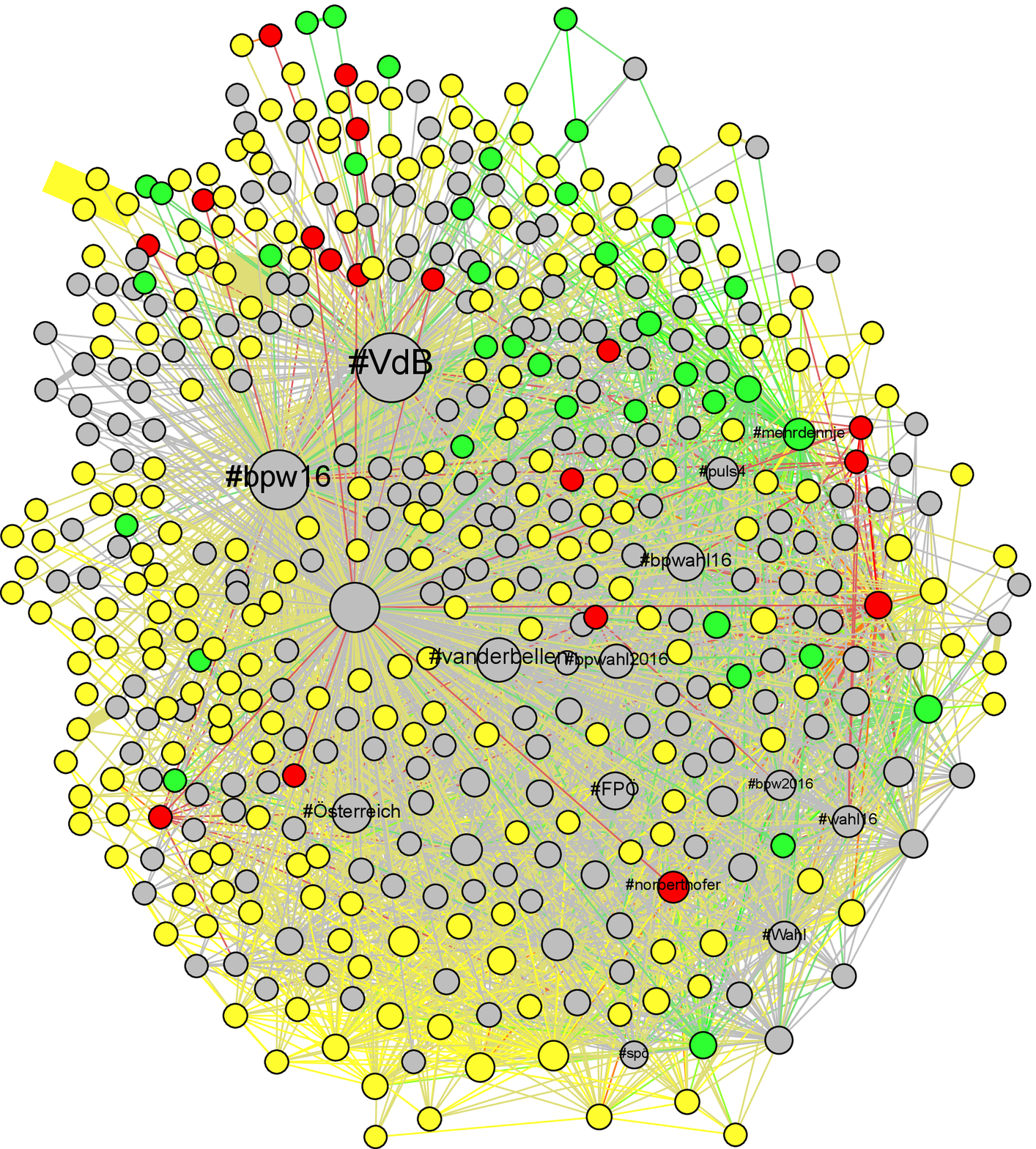}
      \caption{German ego network of Van der Bellen}
			\label{fig:de_egonet_vdb}
			\end{figure} 
      \end{minipage}
      \hspace{0.05\linewidth}
      \begin{minipage}{0.45\linewidth}
         \begin{figure}[H]
       \includegraphics[scale=0.3]{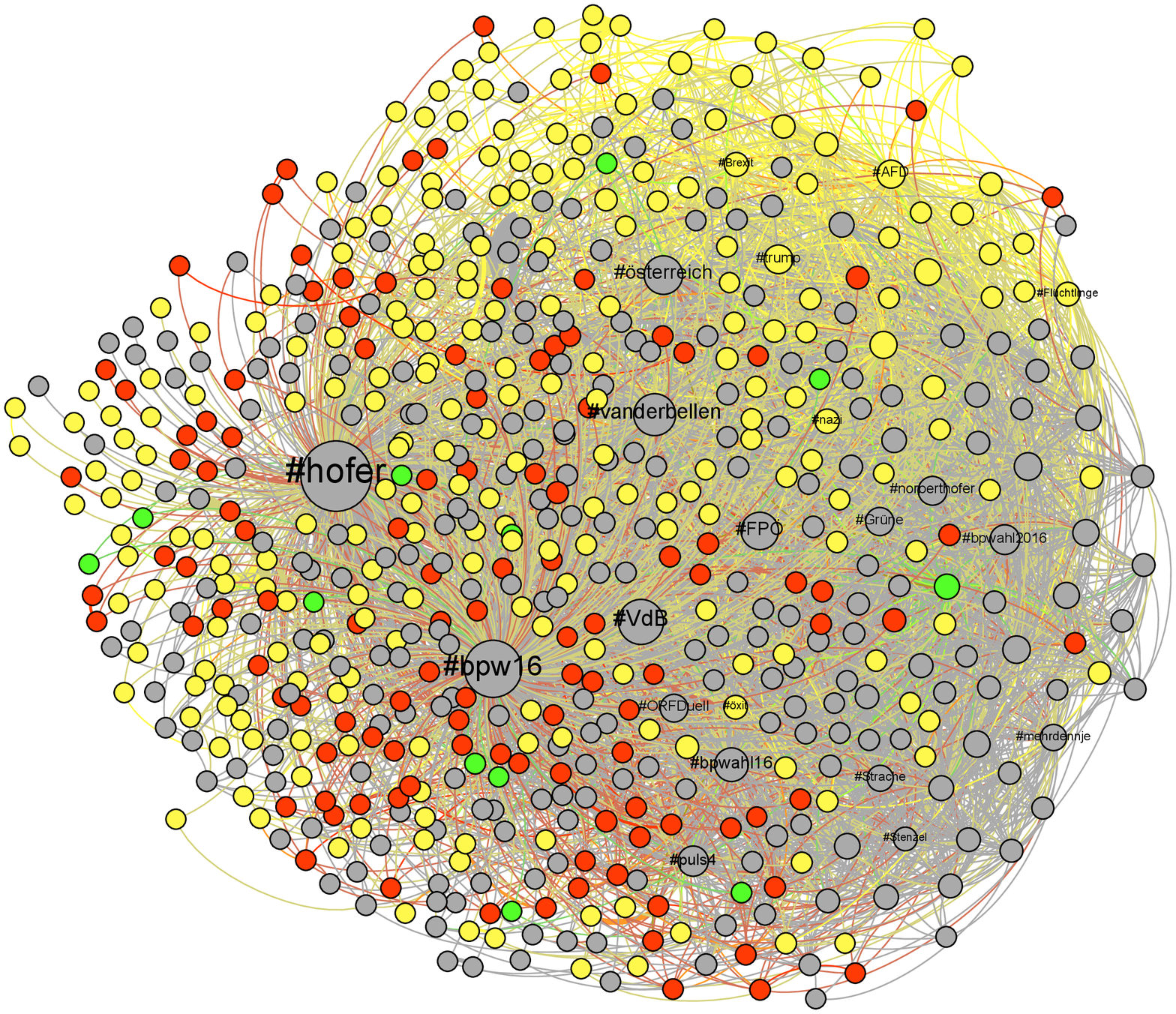}
      \caption{German ego network of Hofer}
			\label{fig:de_egonet_hofer}
			\end{figure}
      \end{minipage}
  \end{minipage}


	\begin{minipage}{\linewidth}
      \centering
      \begin{minipage}{0.45\linewidth}
         \begin{figure}[H]
       \includegraphics[scale=0.4]{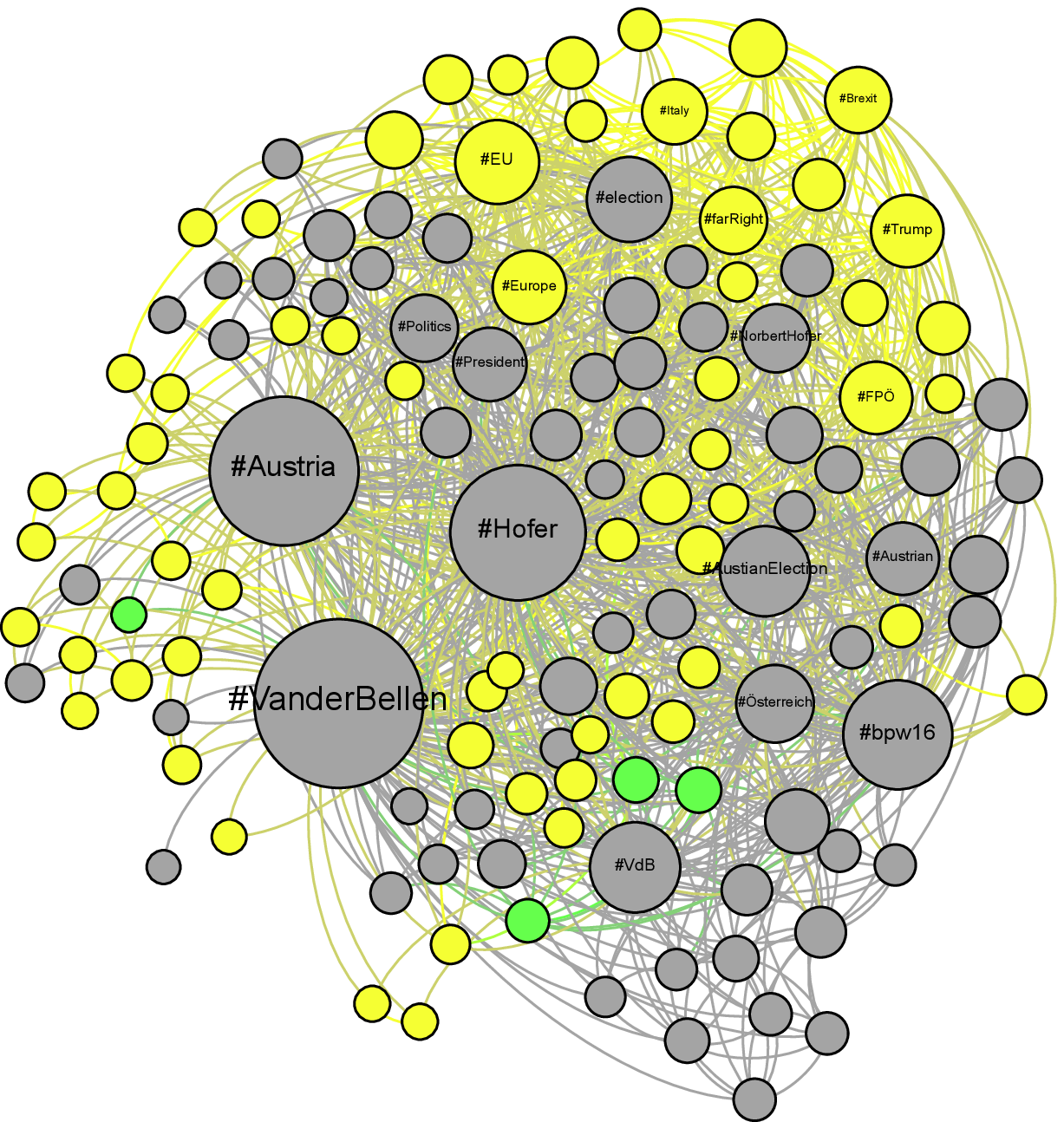}
      \caption{English ego network of Van der Bellen.}
			\label{fig:en_ego_vdb}
			\end{figure}
      \end{minipage}
      \hspace{0.05\linewidth}
      \begin{minipage}{0.45\linewidth}
          \begin{figure}[H]
      \includegraphics[scale=0.4]{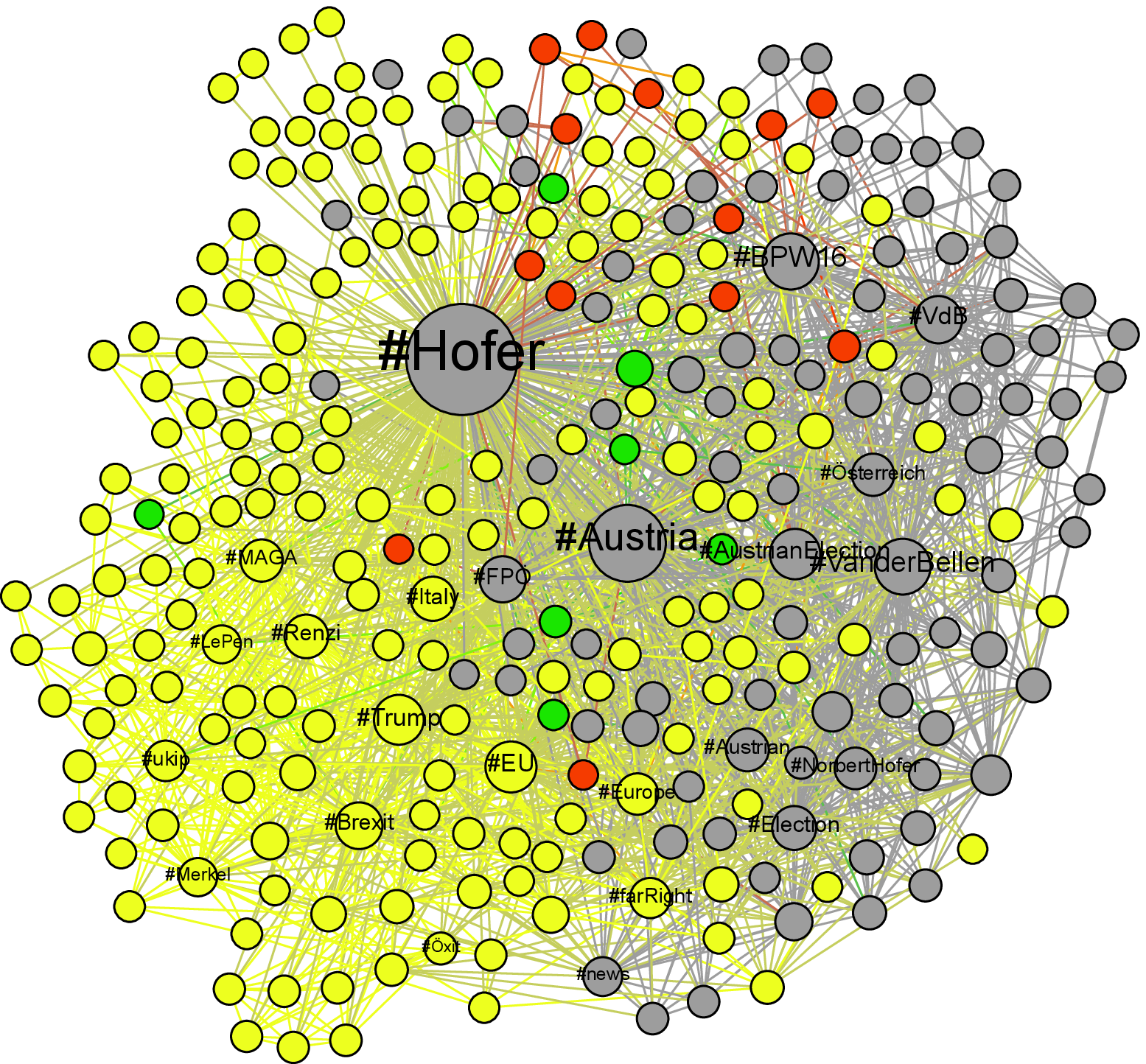}
      \caption{English ego network of Hofer.}
			\label{fig:en_ego_hofer}
			\end{figure}
      \end{minipage}
  \end{minipage}
	
	\bigskip
	
Figures \ref{fig:de_egonet_vdb}, \ref{fig:de_egonet_hofer}, \ref{fig:en_ego_vdb}, and \ref{fig:en_ego_hofer}, show that there is a considerable difference in the way Twitter users refer to each candidate in their tweets. Both, the German and English hashtag ego-networks of Norbert Hofer exhibit more vertices (hashtags) that directly refer to the candidate in a negative context (e.g.\ \textit{\#NoToHofer}) as compared to the ego-networks of Alexander Van der Bellen (see also Table \ref{tab:SummaryOfTheHashtagsCategories}). On the other hand, hashtags that put a candidate in a positive light are used more often in Van der Bellen's ego-networks, as compared to the ones of Norbert Hofer. In Table \ref{tab:SummaryOfTheHashtagsCategories}, we provide a summary of the relative sizes of the hashtag categories, with the maximum value highlighted in bold respectively.

\begin{table}[ht]
	\centering
		\begin{tabular}{l | l | l | l | l}
			\toprule
			& \textbf{Supporting (\%)} & \textbf{Against (\%)} & \textbf{General (\%)} & \textbf{Important topics (\%)} \\
			\midrule
			VDB (de) & \textbf{7.68} & 3.73 & 43.15 & 45.44 \\
			VDB (en) & 3.06 & 0 & \textbf{50.38} & 46.56    \\
			NH (de) & 1.93 & \textbf{18.49} & 41.64 & 37.94 \\
			NH (en) & 2.39 & 4.44 & 32.42 & \textbf{60.75} \\
			\bottomrule
		\end{tabular}
	\caption{Summary of the hashtag categories.}
	\label{tab:SummaryOfTheHashtagsCategories}
\end{table}
	
Although the hashtag ego-networks provide insights into the topics people associate with the candidates, this information alone is not sufficient for deducing the users' general opinion towards each candidate. Thus, in order to get more fine-grained evidence of whether a candidate has been supported by the Twitter users, we also performed an in-depth analysis of the degree-distributions as well as the textual cues and emoticons in the corresponding tweets (see Section \ref{sec:rq22}).

\subsubsection{Communication among the candidates' followers} \label{sec:rq22}

In addition to the ego-networks, we derived the communication network of the candidates' followers\footnote{Note that to preserve the anonymity of Twitter users, we plot the vertices unlabeled in all figures.}. The communication network is a directed network between pairs of usernames that exchanged messages by using the \textit{@} symbol. In particular, a directed edge between two vertices means that the start-vertex of the respective edge sent a Twitter message to the end-vertex of the edge.

\subsubsection*{Overall German language communication network}

The overall German language communication network (including non-followers, i.e.\ Twitter users who participated in the corresponding Twitter discussion but do not follow either candidate) includes 19669 vertices and 58883 edges distributed over 596 connected components. To examine the degree-distribution for the German language communication network, we first derived separate data-sets for the in-degree and the out-degree distribution respectively. Next, we 
removed the vertices having an in-degree or an out-degree of zero\footnote{Remember that the PDF and PMF of the continuous and discrete power-law distributions are defined for $x_{min} > 0$ only. The same goes for the PDFs of the lognormal and exponential distributions (see also the discussion of the model-fitting procedure in this section).}. Such vertices represent users who either did not directly communicate with another Twitter user but just posted tweets that didn't include the \textit{@} symbol (out-degree=0), or users with whom no other Twitter user in our data-set communicated (in-degree=0). In this context, it is worth mentioning that we found a comparatively high number of users that had an in-degree of zero (14272, which corresponds to 72.56\% of all users in our data-set), while a much smaller number of users had an out-degree of zero (2655). 

We then applied a model fitting procedure to the in-degree and out-degree data-sets in order to determine the "best" fit for each of them. In particular, we applied the statistical framework presented by Clauset et al.\ \cite{Clauset2009} and all computations have been performed with the poweRlaw package \cite{powerlaw-package-joss}. Thus, following Clauset et al., we first derived parameter estimates that fit our data to a discrete power-law, lognormal, exponential, as well as a Poisson distribution. For example, for the power-law distribution we had to estimate the lower-bound $x_{min}$ as well as the scaling parameter $\alpha$. The parameter estimation followed the procedure described in \cite{Clauset2009}. 

\begin{figure}[htbp]
    {\centering \resizebox*{0.8\columnwidth}{!}{
        \includegraphics{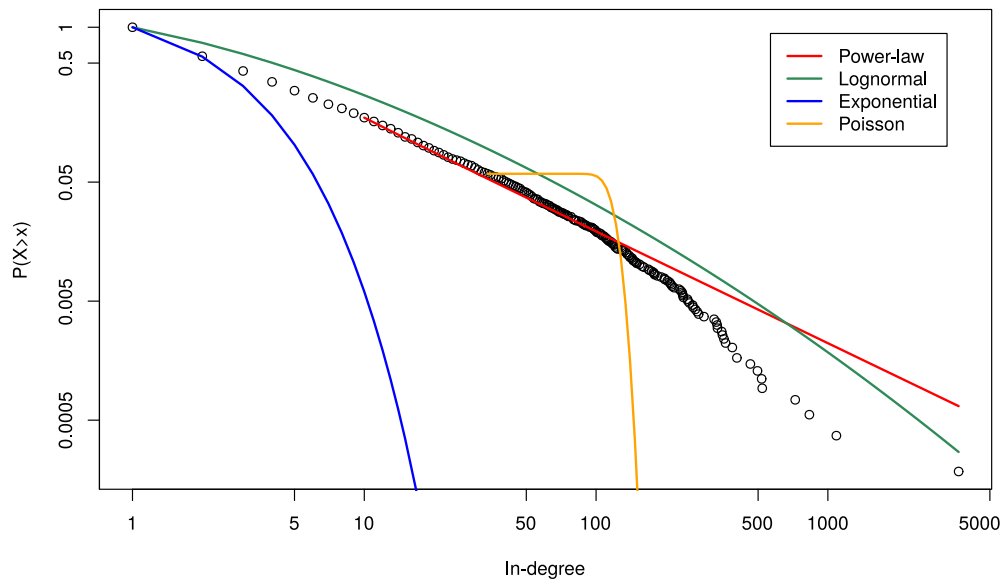}} \par}
    \caption{Best fits for the in-degree distribution.}\label{fig:de_innet_communi_model_estimate}
\end{figure}

\begin{figure}[htbp]
    {\centering \resizebox*{0.8\columnwidth}{!}{
        \includegraphics{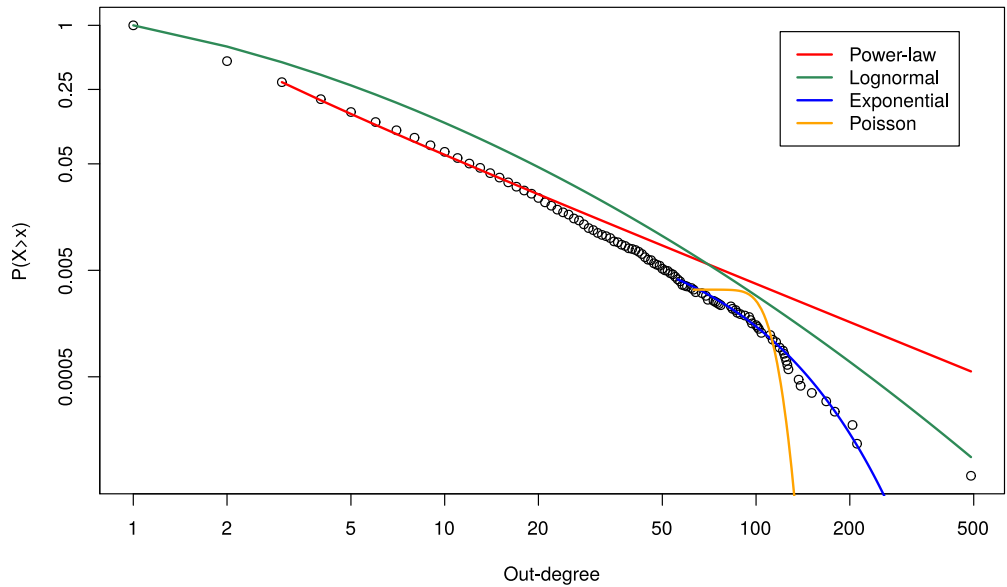}} \par}
    \caption{Best fits for the out-degree distribution.}\label{fig:de_outnet_communi_model_estimate}
\end{figure}

Figures \ref{fig:de_innet_communi_model_estimate} and \ref{fig:de_outnet_communi_model_estimate} show the in-degree and out-degree distributions respectively, together with the four model estimates\footnote{In particular, the figures show the complementary cumulative distribution functions (CCDF) of the estimated models (power-law, lognormal, exponential, Poisson) on a log-log plot.}. Figure \ref{fig:de_innet_communi_model_estimate} also shows a number of outliers which have a significantly higher in-degree than other vertices (users). In particular, the outlier to the far right is the user \textit{@vanderbellen}, who has been talked to much more frequently than all the other users in the data-set (with an in-degree of 3665). Although Norbert Hofer (@norbertghofer) was talked to less frequently (in-degree of 1088), he is still the Twitter user with the second highest in-degree in this communication network.

A visual fitting of statistical models to the true distribution of a data-set is error-prone, though. Moreover, while we can always find some parameter estimates that provide a fit for each of the four models (power-law, lognormal, exponential, Poisson) to a particular data-set, these parameter estimates neither tell us whether a particular fit is good, nor which of the models provides a better fit to the true distribution from which our data was drawn. 

After deriving the parameter estimates, we therefore performed goodness-of-fit tests for the models. In particular, we performed 5000 simulations (i.e.\ generated 5000 synthetic data-sets) to compute a goodness-of-fit p-value for each of the four models. In addition to the goodness-of-fit tests, we directly compared the four models via a likelihood ratio test. For the direct comparisons, we used Vuong's test \cite{Vuong1989} in order to determine if the sign of the respective log-likelihood ratio (LR) is statistically significant (for details of the procedures see \cite{Clauset2009}).

The goodness-of-fit tests for the in-degree distribution resulted in a p-value of 0.1362 for the lognormal and p-values of zero (or approx.\ zero) for all other models, meaning that the lognormal is a plausible hypothesis for the data, while the other hypotheses (power-law, exponential, Poisson) can be rejected. This result was confirmed by the direct model comparisons via the likelihood ratio tests. In particular, for all direct comparisons the log-likelihood ratios indicated that the estimated lognormal model is favored over each of the respective alternatives (power-law, exponential, Poisson)\footnote{Vuong's test resulted in statistically significant p-values with $p \le 0.031$ for each of the direct comparisons.} Thus, according to our model fitting procedure, the lognormal distribution clearly provides the best fit to our data\footnote{The tests we performed cannot tell us if another (heavy-tailed) distribution would provide an even better fit, though.}.

For the out-degree distribution, the situation is not as unambiguous, unfortunately. The goodness-of-fit tests for the out-degree distribution resulted in a p-value of 0.116 for the exponential, a p-value of 0.0142 for the lognormal, and p-values of zero for the power-law and Poisson models. However, the fit for the exponential model has an $x_{min}$ value of 56, meaning that it only fits to the far right tail of the data-set. Moreover, the exponential fit does not explain extreme outliers in the data, such as the vertex with out-degree 490 (see also Figure \ref{fig:de_outnet_communi_model_estimate}). Thus, we again performed direct model comparisons via likelihood ratio tests. Here, the results have been similar to the ones we got for the in-degree distribution. For the out-degree distribution, all log-likelihood ratios again indicated that the lognormal is favored over each of the respective alternatives (power-law, exponential, Poisson)\footnote{Vuong's test resulted in statistically significant p-values equal to zero or approx.\ zero for each of the direct comparisons.}. Moreover, the direct comparisons also indicated that the power-law model is a better fit than the exponential or the Poisson models. Thus, based on the discussion above, none of the models we tested clearly provides the best fit. However, the results of the direct comparisons (likelihood ratio tests) at least indicate that there is a good chance the true distribution of the out-degrees has a heavy-tail.

\subsubsection*{German language communication network of followers}

In order to gain more insight into the communication patterns among the candidates' followers, we also analyzed the ``\textit{follows}'' relation between Twitter users and each presidential candidate. For this part of the analysis, we therefore excluded the remaining vertices (non-followers) from the network (see Figure \ref{fig:en_net_communi}). In particular, each vertex was assigned to one of three categories: followers of Van der Bellen (green), followers of Norbert Hofer (blue), and followers of both candidates (yellow). In this context, it is important to mention that the ``follows'' relation is established between a pair of vertices if a user follows another user (in our data-set one of the presidential candidates) on Twitter. However, the ``follows'' relation does \emph{not} necessarily imply that a user agrees with the candidate's values and ideology. In particular, Twitter users may follow an opposing candidate in order to get more insight into their campaign.

\begin{figure}[H]
   \includegraphics[scale=0.15]{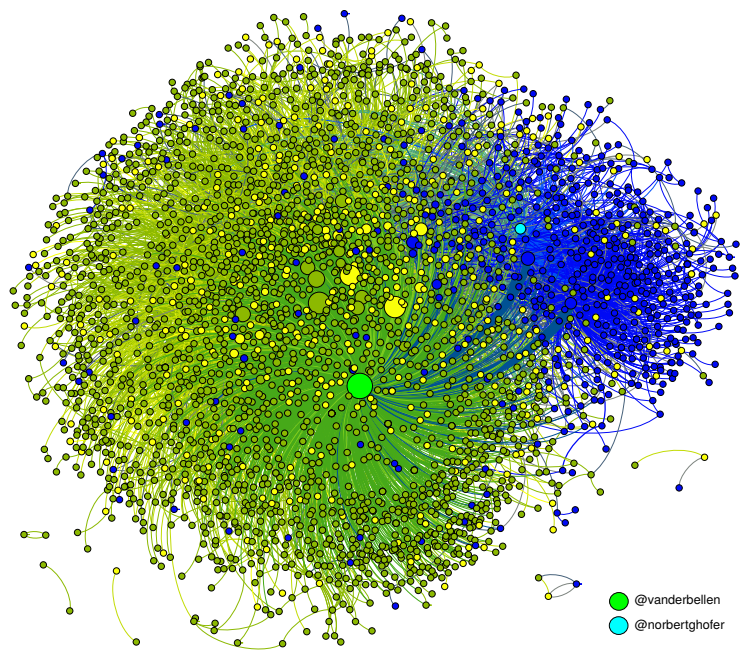}
   \caption{German language communication network of followers. Green vertices are followers of Van der Bellen, blue vertices are followers of Norbert Hofer, and yellow vertices follow both presidential candidates.}
	\label{fig:de_net_communi}
\end{figure}

The German language communication network of followers consists of 2973 vertices and 12278 edges (see Figure \ref{fig:de_net_communi}). The average vertex degree is 4.13 and the number of connected components is 6. In particular, the network contains a giant component including the majority of the vertices as well as five smaller components. 
To measure the influence of the presidential candidates beyond their vertex degree, we applied the betweenness centrality measure\footnote{Note that, because of the nature of the communication network we derived, other centrality measures, such as Eigenvector or PageRank centrality, would not contribute to a significantly improved understanding of the real-world relations between the vertices in this network (for a detailed discussion of different centrality measures see, e.g., \cite{Newman2010, social-and-economic-networks-book, graph-based-social-media-analysis-book, complex-social-networks-book}). In particular, for this network Eigenvector or PageRank centrality would not contribute much beyond degree centrality. Thus, we chose to include a path-based centrality measure (here: betweenness centrality) in addition to degree centrality.}.  The betweeness centrality of Van der Bellen is 634836.01, while Hofer's betweenness centrality is 144046.31, indicating a considerably larger popularity and influence of Van der Bellen's Twitter account with respect to it's degree (see discussion above and below) and betweeness centrality. 

Moreover, the communication network shows a clear distinction of the communication patterns between the followers of Norbert Hofer (blue) and Van der Bellen (green). In particular, the followers of the two opposing candidates tend to form clusters and talk within their group. However, the German communication network also includes a number of individual blue vertices (followers of Norbert Hofer) who engage in a communication with the followers of Van der Bellen. 

Since the network of followers is a digraph, we again distinguish the in-degree and the out-degree distribution (see Figures \ref{fig:de_net_communi_indegree_dist} and \ref{fig:de_net_communi_outdegree_dist}). 
In this network, \textit{@vanderbellen} is the vertex with the highest in-degree (1290) (i.e.\ compared to all other vertices in the network, Van der Bellen was the one who has been directly addressed most often). Norbert Hofer also belongs to the vertices with a high in-degree (496), however, significantly lower than Van der Bellen. Thus, our data-set shows signs of a so-called \textit{broadcast in-hub network} \cite{Smith2014}, which is a network with a vertex (in our case \textit{@vanderbellen}) with many incoming edges, while the vertices pointing to the hub are themselves not tightly connected and may form only smaller subgroups\footnote{After extracting the ego-network of the \textit{@vanderbellen} vertex, the resulting network density was $\approx 0.2$\%, which indicates that users talking to \textit{@vanderbellen} tend not to communicate amongst each other. The ego-network of Hofer also exhibits a \textit{broadcast in-hub network} (\textit{@norbertghofer}) and a network density of $\approx 0.8$\%.}.

Figure \ref{fig:de_net_communi_outdegree_dist} shows the out-degree distribution of the German language communication network of followers. The plot shows that both presidential candidates communicated comparatively less often than some other users participating in the Twitter discussion. The most active user (in terms of communication with other users) in the network sent more than 500 tweets, while \textit{@vanderbellen} and \textit{@norbertghofer} sent less than 40 tweets directly to one of the users that officially follow the candidates\footnote{A word of caution is in order here: While the candidates appear to be engaging rarely in a one-to-one communication, it is important to note that the actual one-to-one count exceeds the ones presented in Figure \ref{fig:de_net_communi_outdegree_dist}. This is due to the fact that we only show an extract of the network in which we filtered out the vertices that do not follow the candidates on Twitter. For example, many European politicians congratulated Van der Bellen on his victory and Van der Bellen responded to those tweets. However, such one-to-one messages are not included in Figure \ref{fig:de_net_communi} because there is no follower-relationship between the respective user accounts.}. In Section \ref{sec:bots}, we provide an additional analysis on the meaning of high numbers of outgoing messages from particular users and contrast them to the behavior of bots that we identified in our data-set.

	\begin{minipage}{\linewidth}
      \centering
      \begin{minipage}{0.45\linewidth}
         \begin{figure}[H]
					 \includegraphics[scale=0.4]{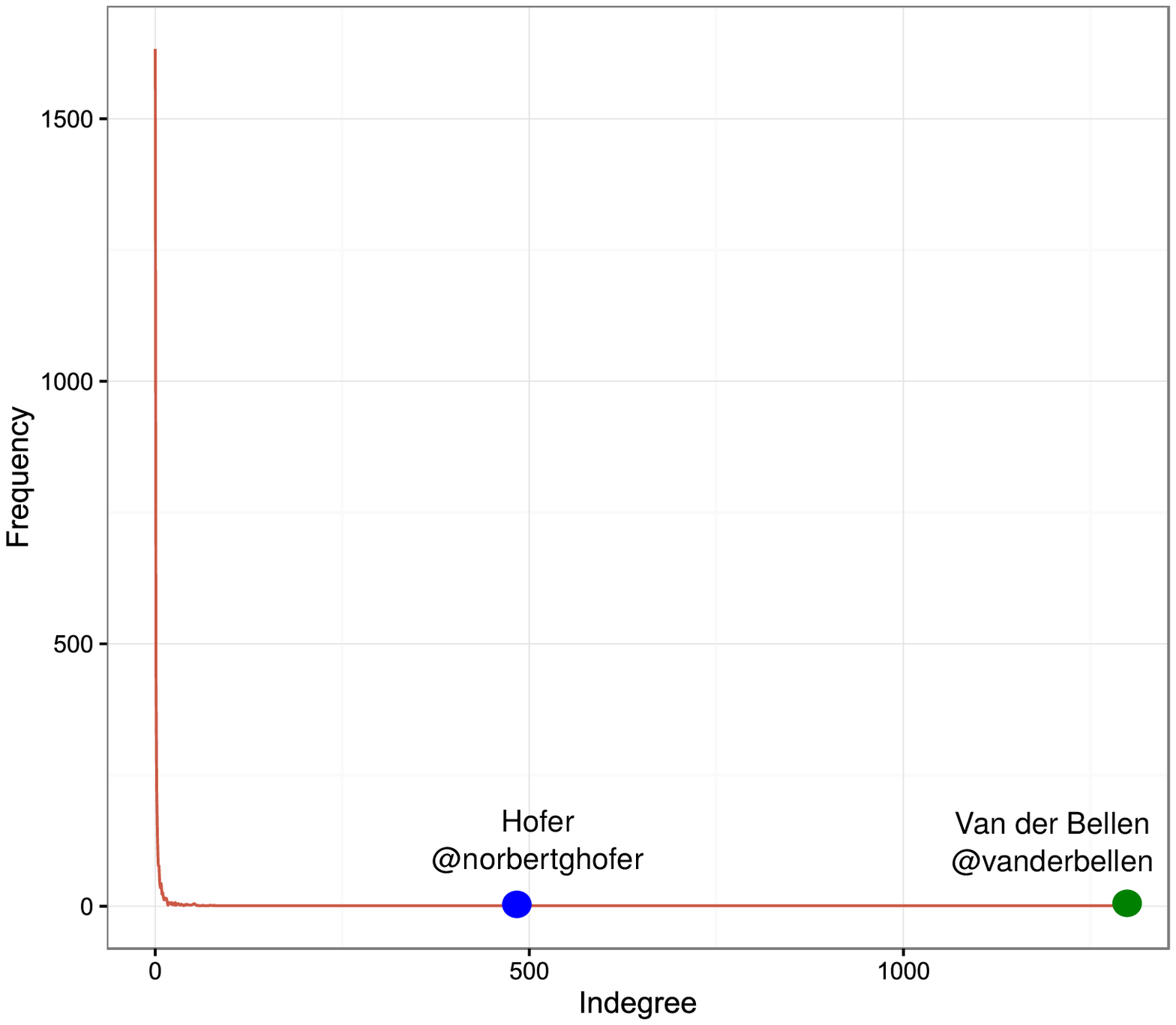}
					 \caption{In-degree distribution of the German communication network.}
					\label{fig:de_net_communi_indegree_dist}
				\end{figure}
      \end{minipage}
      \hspace{0.05\linewidth}
      \begin{minipage}{0.45\linewidth}
         \begin{figure}[H]
						 \includegraphics[scale=0.4]{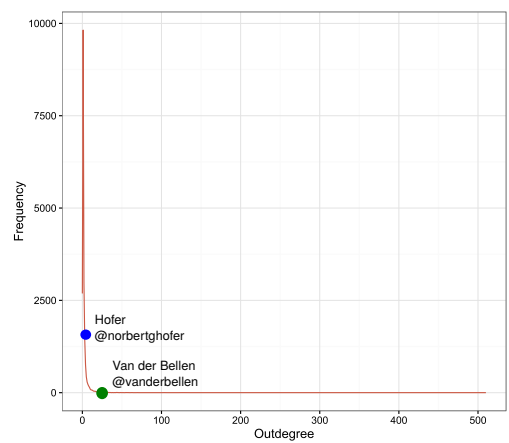}
						 \caption{Out-degree distribution of the German communication network.}
						\label{fig:de_net_communi_outdegree_dist}
					\end{figure}
      \end{minipage}
  \end{minipage}
	

\subsubsection*{English language communication network}

\begin{figure}[H]
   \includegraphics[scale=0.12]{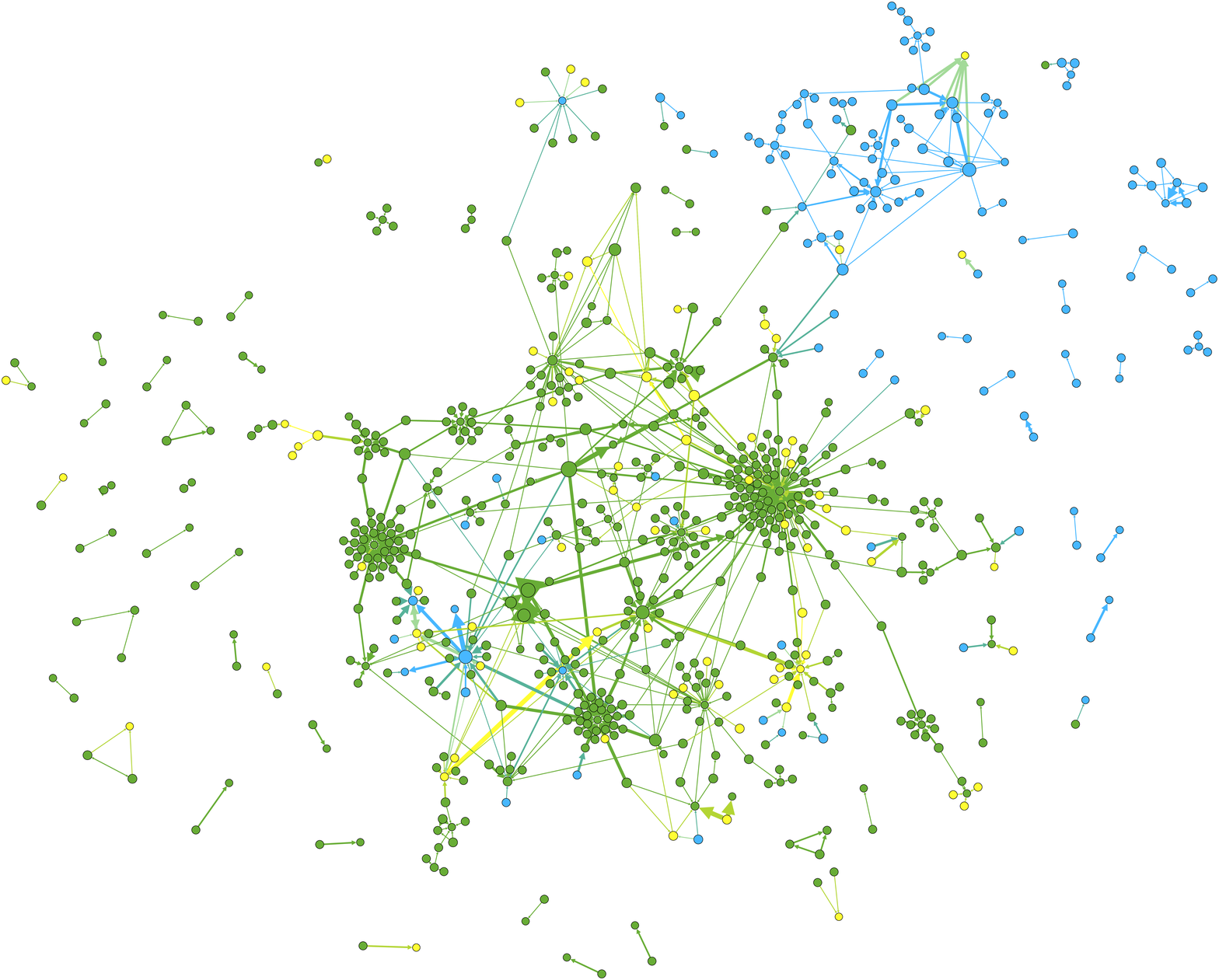}
   \caption{Excerpt of the English language communication network.}
	\label{fig:en_net_communi}
\end{figure}

The communication network resulting from the English language data-set is a directed network with 69588 vertices, 89862 edges, and 2072 connected components.  Figure \ref{fig:en_net_communi} shows a subgraph including 733 vertices, 859 edges, 54 connected components, and an average vertex degree of 1.17. Thus, this particular subgraph includes 1.05\% of all vertices. We visualized this subgraph, because it exemplifies the user polarization and a tendency of the candidates' followers to communicate within the group they belong to. 

\begin{figure}[ht]
	\centering
		\includegraphics[scale=0.65]{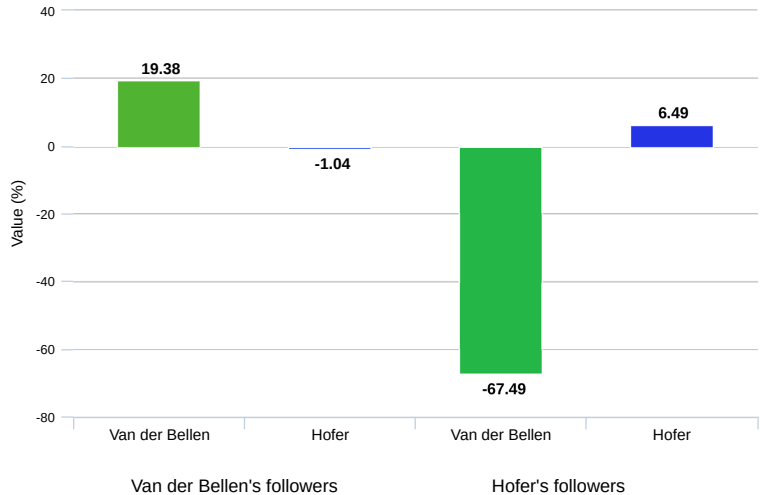}
		\caption{Supporters' opinions about their candidate and his respective rival.}
	\label{fig:de_opinion_opposite_candidate}
\end{figure}

In addition to the German and English language communication networks, we also analyzed the opinion of the followers in both communication networks regarding their candidate and his respective opponent. We do this by first separating the data-set into two groups of followers, one for each candidate. For each group we used regular expressions to obtain tweets that mention each candidate individually and extract the sentiment scores as assigned by SentiStrength \cite{Thelwall2010}. If both candidates were mentioned in a tweet, we extracted the corresponding tweets and manually classified them as positive or negative for each candidate. For example, a tweet from our data-set \textit{``God knows Norbert Hofer is a Christian and Van der Bellen is Godless''} is classified under positive for Norbert Hofer and negative for Van der Bellen. We stored this information and extracted the sentiment scores assigned by SentiStrength accordingly. As illustrated in Figure \ref{fig:de_opinion_opposite_candidate}, there is a noticeable difference in the way the followers address the candidates. While Van der Bellen's followers predominantly disseminate positive sentiments regarding Van der Bellen, with a comparatively small number of negative sentiments about Norbert Hofer, the followers of Norbert Hofer tweet mostly negatively about Van der Bellen. Such tweets even substantially exceed the positive tweets about Norbert Hofer.

\subsection{Analysis of potential bot activities (RQ3)}\label{sec:bots}

A \emph{social bot} is an autonomous piece of software with the goal to either increase a user's popularity \cite{Ratkiewicz2011}, to influence users \cite{Boshmaf2011}, or intentionally harm users by misleading, destroying reputation, manipulating public opinion \cite{ieee-computer-darpa-twitter-bot-challenge}, and jeopardizing democracy \cite{cacm-rise-of-social-bots}. In the past, bots mainly re-created content (e.g.\ by re-tweeting messages). Thus, the first bot detection techniques primarily dealt with identifying social media accounts that do not follow human-like behavioral patterns. However, recently bots have become more refined and are able to mimic people. For example by actively participating in conversations, following trendy topics, or following users, which again poses significant challenges in detecting them \cite{ieee-computer-darpa-twitter-bot-challenge}. 

It has been estimated that in 2009 bots generated about 24\% of the content on Twitter  \cite{Tsvetkova2016}, and different analyses reported that the social media discussions on both, the 2016 Brexit referendum as well as the 2016 US presidential elections have been strongly influenced by bots (see, e.g., \cite{social-bots-distort-2016-us-election, Howard2016, Kollanyi2016}). Thus, we applied two bot detection techniques to analyze whether bots have been involved in the discussion about 2016 Austrian presidential elections. Because the official language in Austria is German, we focused our analysis on the German language data-set. In our analysis, we extracted and examined 22450 unique user accounts for signs of bot behavior\footnote{Since users may delete their accounts (i.e.\ Twitter accounts may not exist anymore at the time of the analysis) or the information about a Twitter user might be insufficient for the analysis, we were not able to generate bot scores for 2.4\% of the user accounts in our analysis.}. Our bot analysis combined two approaches. In particular, we used the \textit{BotOrNot} Python API \cite{Davis2016} and applied a content generation count. Moreover, we explicitly searched for the word \textit{bot} in Twitter user-names to identify self-disclosed bots (as suggested in \cite{Howard2016, Kollanyi2016}). 

The \emph{BotOrNot} API provided us with scores in the closed interval [0-1], where 0 indicates that the account is not a bot and 1 that it is a bot with a 100\% probability. Since \emph{BotOrNot} scores range from 0 to 1, we split the scores into three categories:

\begin{itemize}[noitemsep]
\item accounts with scores between 0 and 0.5 were marked as human accounts,
\item accounts with scores between 0.51 and 0.90 were marked potential bots, 
\item accounts with scores between 0.91 and 1 were marked as bots. 
\end{itemize}

Out of 22450 accounts, \emph{BotOrNot} identified the majority (20645, 91.96\%) as human accounts, while 1117 (4.98\%) accounts were marked as potential bots, and 148 (0.66\%) accounts were identified as bots with a probability between 91.00-100\%.

Furthermore, by applying the content generation count approach, we marked an account as a bot if it had a tweet count of 50 tweets/day or more for at least three days during the extraction period, or if the username contained the word \textit{bot} \cite{Howard2016, Kollanyi2016}. This procedure resulted in the identification of 20 (potential) bot accounts (0.09\% of the users). A comparison with the respective \emph{BotOrNot} results shows that \emph{BotOrNot} assigned 16 of those 20 to the \textit{human account} category and 4 to the \textit{potential bot} category.

While other studies found evidence of significant Twitter bot activity in recent events such as the 2016 US presidential elections or the 2016 Brexit referendum
\cite{social-bots-distort-2016-us-election, Howard2016, Kollanyi2016}, our results indicate that this was not the case during the 2016 Austrian presidential elections 2016 (see Table \ref{tab:ComparisonOfBotActivity}).  

\begin{table}[ht]
	\centering
		\begin{tabular}{ccccc}
			\toprule
			&\textit{US elections} & \textit{Brexit} & \textit{AT elections (BD1)} & \textit{AT elections (BD2)}\\
			\midrule
			Generated content & 27.2\% & $\approx$ 30\% & 1.02\% & 5.29\% \\
			RT received				& - 		 & - 							& 0.99\% & 0.96\% \\ 
			Like count 				& - 	 	 & - 							& 0.57\% & 2.34\% \\
			RT generated			& -			 & -							& 1.07\% & 5.36\% \\
			VDB follower			& -			 & -							& 0.91\% & 0.22\% \\
			NH follower 			& - 		 & -							& 0.94\% & 0.11\% \\
			\bottomrule
		\end{tabular}
	\caption{Comparison of the Twitter bot activity during important political events in 2016\protect\footnote{In the table, \textit{BD1} refers to bot detection via \textit{BotOrNot}, and \textit{BD2} refers to the technique that relies on the content generation count and self-disclosed bots.}.}
	\label{tab:ComparisonOfBotActivity}
\end{table}

\begin{figure}[ht]
	\centering
		\includegraphics[scale=0.9]{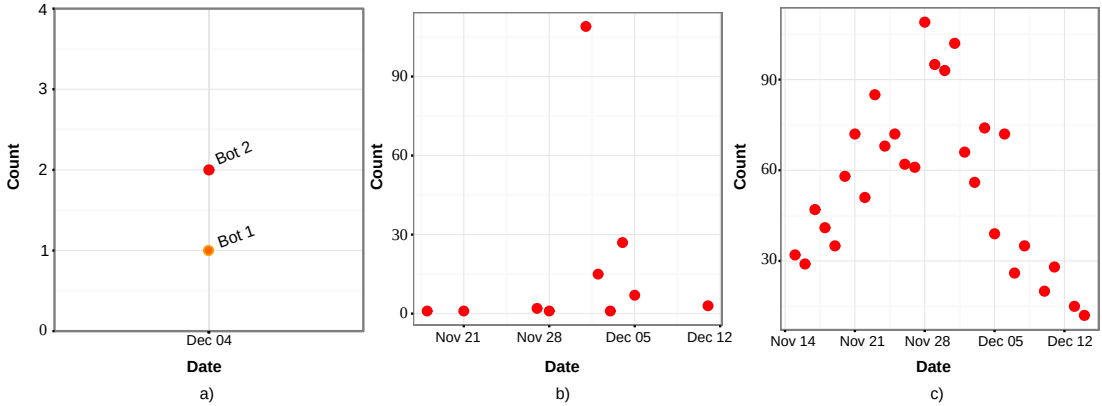}
	\caption{Tweeting behavior of prospective bot accounts.}
	\label{fig:bots_summary}
\end{figure}

Figure \ref{fig:bots_summary} shows daily tweeting activities of the four accounts marked as bots by both bot detection approaches.  Figure \ref{fig:bots_summary}a) shows the tweeting behavior of two self-disclosed bots, where bot\_1 sent one tweet and bot\_2 sent two tweets on election day. Figure \ref{fig:bots_summary}b) shows the more human-like behavior of another self-disclosed bot with a smaller number of daily tweets prior the election day, an increase shortly before and during the election day, with one day of extreme tweeting activity (an outlier). Finally, Figure \ref{fig:bots_summary}c) shows the tweeting activity of a highly active bot account. It is important to note that while observing the tweeting activity of the four prospective bots, one can observe three distinctive types of tweeting, i.e. in contrast to our expectations, the four prospective bots did not exhibit a similar (or even uniform) tweeting pattern.

\section{Discussion}\label{sec:discussion}

Over the past decade, social media platforms have become an important channel for people to search information, participate in discussions, and share their opinion. As a consequence, the scientific community came up with an increasing number of studies that report on different aspects of social media user behavior. These aspects generally cover the quantitative analysis of information sharing (e.g.\ tweet and re-tweet count, see, e.g., \cite{twitter-use-by-us-congress, towards-more-systematic-twitter-analysis-metrics}), sentiment analysis of the messages being spread (see, e.g., \cite{emotional-valence-shifts-asonam-2017, word-emotion-lexicons-snams-2017, Diaz2012}), as well as network analysis of communication relations (via \textit{@username}) or hashtag networks (see, e.g., \cite{power-of-social-media-analytics, Fornacciari2015}). Even though studying each of these aspects of user behavior in isolation may already provide interesting insights, a more comprehensive portrayal of user behavior can only be obtained if the aforementioned aspects are combined. 

The analysis performed for this paper combines methods from sentiment analysis, network analysis, and bot detection. It thereby complements existing studies on other elections and provides a generic approach that can be used for analyzing future social media events (see also Sections \ref{sec:approach} and \ref{sec:generic-approach}).

In our analysis of the 2016 Austrian presidential elections, we found a clear pattern which shows that emotional tweets (negative as well as positive) are re-tweeted, replied to, and liked more often than the neutral ones. We can thereby confirm the findings from \cite{Stieglitz2013} which reported on similar findings for microblogs. An explanation for such a user behavior is that emotionally charged tweets also trigger emotions in other users. For example, an expression of gratitude (identified in both candidates' data-sets), a message of warning against the FPÖ candidate sent by a Holocaust survivor, and a message of sadness after loosing the elections generally received a high number of reactions (likes, replies, and re-tweets). 

In addition, we found that occasionally tweets with a neutral sentiment score may also be quite influential in terms of the reactions caused in other users. For example, Twitter users highly liked, re-tweeted, and replied to tweets that carried an invitation for the people to vote. A possible explanation is that such tweets convey an implicit emotion (here: anticipation) which causes a feeling of urgency and importance (e.g.\ one the respective tweets said \textit{``today every single vote counts''}). 

In order to analyze the structural properties of the communication network, we derived hashtag ego-networks which helped significantly to gain further insight into which topics users associate with the candidates. In particular, we found that hashtags related to important topics are predominant in Norbert Hofer's English language ego-network (60.75\%). Such hashtags often carry a controversial note -- e.g. ``make Austria great again'', ``Trump'', ``Öxit'', ``Brexit'', etc. In part, such an ego-network might be attributed to the influence of other media sources (radio shows, online news, etc.), which put Hofer in correlation with the controversial topics (see, e.g., \cite{Oxit2016}). 
Here it is important to note that one might also approach the task of studying associated topics by constructing a network of terms used in the tweets. For example, \cite{Song2014} studied the topics associated with each presidential candidate during the 2012 Korean presidential elections and re-constructed a network of terms that appeared in the same tweet. In our case, we found that Van der Bellen was associated with the topics of respecting human rights and welcoming refugees, to name a few, while his opponent was associated with loving the homeland and protecting the borders. This confirms that the candidates' statements from the TV discussions directly found their way into the social media discussion on Twitter, even though neither of the candidates directly posted a tweet with a corresponding message. 

With respect to the influence of more traditional media (i.e.\ non-social-media channels such as newspapers or TV channels), we also examined cases of negative campaigning. In our analysis of negative campaigning, we made use of reputable Austrian media (newspapers and TV channels) who published evidence that either confirmed or rejected negative rumors. We used these sources to obtain a list of keywords to find occurrences of misinformation and negative campaigning in our English and German language data-sets. In particular, we performed a time-series analysis, follower analysis, and content analysis, and were able to determine the impact of such tweets on the Twitter discourse. 

While aligning our findings with existing rumor theories, we found evidence that complies with Rosnow's theory that rumors propagate because of the people's tendency to clarify uncertain events \cite{Rosnow1991}. Another study which complements our findings \cite{Pattie2011} discussed potential risks to a person spreading misinformation. Our findings confirm that misinformation may lead to negative consequences for the spreader and backfire against him/her. In particular, the accusation of Van der Bellen being a spy is a good example where Van der Bellen's followers participated in spreading the misinformation over Twitter. In this particular part of our analysis, the importance of using an integrated data analysis approach is evident. As an example of how a combination of analysis methods help to correctly classify user behavior, we refer the reader to the ironic reaction by one of Van der Bellen's followers (cf.\ Section \ref{sec:rq1}).
Without the combined analysis approach that we applied to the 343645 unique tweets in our data-set, such a message would have been falsely classified as misinformation or a tweet that goes in disfavor of Van der Bellen. It is, however, important to correctly classify such messages since irony or sarcasm are commonplace in social media messages, and a wrong classification of such messages may lead to false conclusions when interpreting social media user behavior.

\subsection{Systematic approach for studying social media events} \label{sec:generic-approach}

\begin{figure}[htbp]
	\centering
		\includegraphics[scale=0.7]{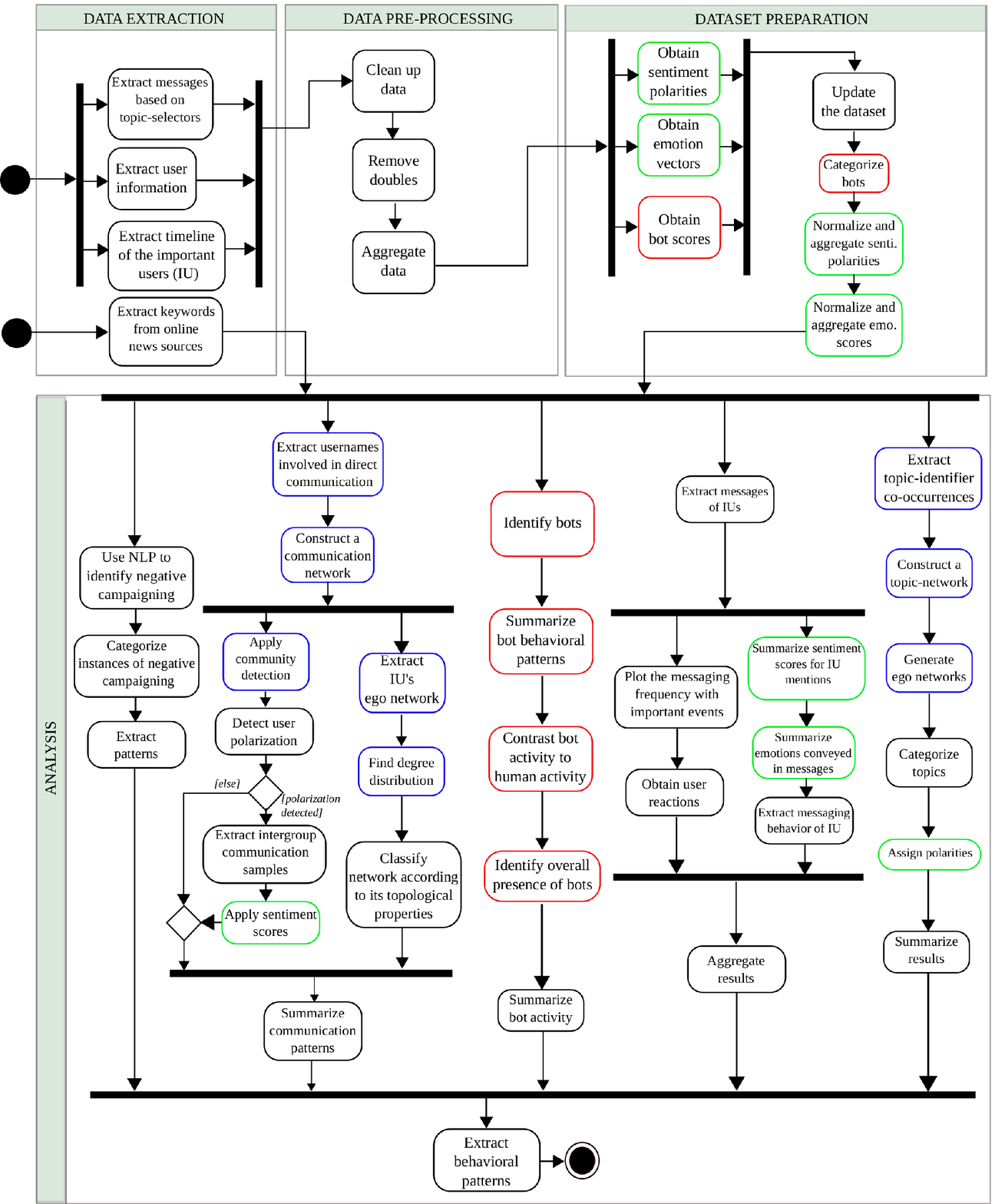}
	\caption{Generic approach for analyzing social media events\protect\footnote{Actions with a green frame are related to sentiment analysis steps, actions with a blue frame are related to network analysis steps, actions with a red frame are related to bot detection steps.}.}
	\label{fig:method}
\end{figure}

While the case study presented in this paper provides an in-depth analysis of the Twitter discussion surrounding the 2016 Austrian presidential election, 
our approach for integrating sentiment analysis, network analysis, and bot detection is generic and can be applied to a wide variety of social media events (see also Section \ref{sec:approach}). Figure \ref{fig:method} shows an UML activity diagram \cite{uml-superstructure-2.5} that visualizes the approach.

\textbf{Data extraction phase:} As in every data analysis process, a systematic data extraction procedure is the first step for analyzing user behavior. In particular, this includes selecting appropriate tools and getting acquainted with the corresponding APIs. Another important decision is choosing the topic-selectors (e.g.\ hashtags) for identifying and extracting relevant messages (e.g.\ tweets). For our case study we carefully selected appropriate hashtags to focus the data extraction towards relevant tweets (see also Section \ref{sec:method}).  

\textbf{Data pre-processing:} After extracting the raw social media data, it is important to identify and remove doubles. Moreover, we have to clean the data-set from misspellings and uninformative chunks of symbols (e.g.\ HTML tags, URLs, or line breaks) which might interfere with subsequent analysis steps (see also Section \ref{sec:method}). After data cleaning and correction, one often performs first aggregation steps over the data. For example, in our case study the usernames in tweets have been matched to usernames in the list of the candidates' followers. 

\textbf{Data-set preparation:} After the pre-processing steps are done, the data-set is ready for running preparatory steps needed for the user behavior analysis. This typically includes the application of software tools for deriving bot scores, sentiment scores, and emotion vectors from the pre-processed data-set. In particular, for our case study we applied the SentiStrength algorithm \cite{Thelwall2010} to obtain polarity scores and used the NRC emotion-word lexicon \cite{Mohammad13} to identify the emotion vectors associated with the tweets in our data-set. Moreover, we obtained bot scores by using the \textit{BotOrNot} Python API \cite{Davis2016}. 

\pagebreak

\textbf{Data analysis:} Once the data-set is complete (including bot scores, sentiment scores, and emotion vectors), we can start the actual data analysis. In particular, four different types of analyses are performed (see also Figure \ref{fig:method}):

\begin{itemize}
\item a \emph{quantitative data analysis} to summarize the users' communication behavior (e.g.\ the tweeting behavior) in terms of the number of likes, replies, and endorsements (e.g.\ re-tweets); 
\item a \emph{sentiment analysis} to identify sentiment polarities and emotions communicated by social media users (e.g.\ the presidential candidates and their respective followers); 
\item a \emph{network analysis} to study the communication patterns of social media users (e.g.\ via ego-networks of important nodes such as the presidential candidates; and topic-networks, such as hashtag-networks, to determine which topics are related to each other and/or associated with certain users). 
\item a \emph{bot analysis} to identify automated software agents (social bots) that try to influence human users.
\end{itemize}

Moreover, for the analysis of the 2016 presidential election we also applied natural language processing (NLP) techniques \cite{Chowdhury2003,Surabhi2013} to identify occurrences of negative campaigning. 

\subsection{Limitations}

For our study the main restrictions result from the tools we used for data extraction, pre-processing, preparation, and analysis (see Section \ref{sec:approach}). 
In particular, we used the Twitter API to extract publicly available tweets. One significant limitation is an API restriction which only allows for the extraction of tweets that are at most seven days old. Thus, if not planned properly, the data extraction cannot be repeated because the API restricts access to tweets older than a week. Moreover, Twitter explicitly says that not all tweets are indexed or made available by the Twitter API \footnote{https://dev.twitter.com/rest/reference/get/search/tweets}. Thus, even though we performed a systematic procedure where we extracted the new tweets on a daily basis, we cannot rule out the possibility that we missed relevant tweets due to this API restriction. In this context, it is important to mention though that the data-sets we extracted for the two presidential candidates (@vanderbellen and @norbertghofer) are complete and include all tweets that have been sent during the extraction period. We ensured this completeness by checking the tweets extracted via the API and manually added tweets that were omitted by the Twitter API. However, it was infeasible to repeat the same procedure for all tweets (i.e.\ all tweets of the candidates' followers) since that would have meant to manually check several ten-thousand user profiles on a daily basis.

A second limitation comes with the tools that we used for data analysis. In particular, we used the sentiment analysis tool SentiStrength, the NRC emotion-word lexicon, and the \textit{BotOrNot} Python API. Even though the three tools have been used in numerous related studies (see e.g. \cite{Thelwall2010, Davis2016, Mohammad13, Stieglitz2013}), we cannot exclude the possibility that some scores assigned by the tools are not appropriate. Thus, to mitigate such errors, a prior assessment of the tools (e.g.\ by deploying human raters) could improve the overall correctness of the assigned scores (see, e.g., \cite{word-emotion-lexicons-snams-2017}). 

\section{Related work} \label{sec:RelatedWork}

Over the past decade, a number of authors studied the diffusion of tweets in the context of political campaigns.  Many authors focus on a single type of analysis though. In contrast, the study presented in this paper provides a multi-faceted analysis including sentiment analysis, social network analysis, and bot detection (see Section \ref{sec:generic-approach}).

In \cite{Stieglitz2013}, Stieglitz and Dang-Xuan applied the SentiStrength algorithm for sentiment analysis to study the spread of tweets during the German elections in 2012. In particular, they quantified the impact of positive and negative tweets in terms of the re-tweet count and the speed of re-tweeting. In \cite{Diaz2012}, Diaz-Aviles et al.\ studied the public opinion about the presidents of 18 Latin American countries by applying sentiment analysis techniques to Spanish language tweets and short blog posts. To determine how people feel about each president, the authors carried out a part-of-speech tagging to extract the list of nouns and adjectives which they later mapped to a corresponding emotion score in the NRC emotion lexicon. 
Additional non-English language studies have been conducted for the Nigerian presidential elections 2011 \cite{Fink2013}, Indonesian presidential elections \cite{Gemilang2014}, as well as the Bulgarian presidential elections \cite{Smailovic2015}.

Some studies combine sentiment analysis and social network analysis. For example, in \cite{Bermingham2009} Bermingham et al.\ study jihadists' radicalization over social networks. They took a lexicon-based approach to identify sentiment polarities in YouTube comments and combined it with the network aspects of information sharing. In particular, they applied betweenness centrality to identify influential users in the YouTube-sphere, analyzed the network density, and determined the average communication speed. In another study Fornacciari et al.\ \cite{Fornacciari2015} examined the differences in opinions among Twitter communities by reconstructing a follower-followee network of over 60 Twitter channels and assigning sentiment polarity scores to each vertex (user). 

Some studies merely focused on the application of network analysis methods. For example, in \cite{Burgees2012} Burgees and Bruns collected tweets about the 2010 Australian elections containing the \textit{\#ausvotes} hashtag. In particular, they investigated the topics people tweeted about and reconstructed a network of replies. The authors distinguished between a passive (broadcast only) and an interactive user behavior, and identified important users in the network by applying the betweenness centrality measure. In \cite{Song2014}, Song et al.\ applied a latent Dirichlet allocation over a set of tweets to identify a list of topics discussed during the 2012 Korean presidential elections. They examined the occurrences of each topic within a time period and categorized them as a rising (trending) or a falling topic. In addition, they studied the topics related to each presidential candidate and constructed a network of term co-occurrences.     

In the light of recent events (such as the 2016 US presidential elections and the 2016 Brexit referendum), numerous authors reported on the misuse of social media channels aiming to manipulate the voters' opinions (cf.\ \cite{alchemy-of-authenticity-lessons-from-2016-us-campaign, explaining-donald-trump-via-communication-style, Howard2016}). However, such a misuse has already been reported much earlier. For example, in \cite{Ratkiewicz2011} Ratkiewicz et al.\ discuss the issue of political abuse by studying the spread of misinformation. The authors consider mood scores based on Google's Profile of Mood States (calm, alert, sure, vital, kind, happy) (see \cite{Bollen2011}) and basic properties of the hashtag network and user-mention (i.e.\ \textit{@username}) network topology to automatically classify messages as truthful or fake. Another study by Jin et al.\ \cite{Jin2014} investigates the characteristics of the spread of misinformation concerning Ebola by applying epidemiological modeling (see also \cite{Bettencourt2006}). 
Moreover, in \cite{Howard2016} Howard and Kollanyi report that bots have been used to amplify messages by raising their re-tweet count in order to influence the 2016 Brexit referendum. Even though Howard and Kollanyi identified only a small amount of bots participating in discussions about Brexit, those bots created a comparatively large amount of the overall content (about one third of all messages). 
			
\section{Conclusion} \label{sec:conclusion}

In this paper, we presented an analysis of the Twitter discussions surrounding the 2016 Austrian presidential elections. We extracted and analyzed 343645 German and English language Twitter messages that have been posted by the two presidential candidates, their supporters, as well as other Twitter users.   
In particular, we specified, documented, and applied a systematic approach for analyzing social media user behavior (see Sections \ref{sec:method} and \ref{sec:generic-approach}). Our approach is generic and can be applied to perform similar analyses for a wide variety of other social media events. Moreover our study combines sentiment analysis, network analysis, as well as bot detection. Thereby, our paper complements previous approaches and case studies for analyzing  social media user behavior in large real-world events (cf.\ Section \ref{sec:RelatedWork}).

For the 2016 Austrian presidential election, we found that:

\begin{itemize}
\item with respect to our network analysis, the winner of the election (Alexander Van der Bellen) was considerably more popular and influential on Twitter than his opponent;
\item in their attempt to correct misinformation, the followers of Van der Bellen also substantially participated in the spread of exactly that misinformation about him;
\item there was a clear polarization in terms of the sentiments spread by Twitter followers of the two presidential candidates; 
\item the in-degree and out-degree distributions of the underlying communication network are heavy-tailed; 
\item compared to other recent events, such as the 2016 Brexit referendum or the 2016 US presidential elections, only a very small number of bots participated in the Twitter discussion on the 2016 Austrian presidential election.
\end{itemize}
 
In addition to the main findings summarized above, we also found that the two presidential candidates showed a tendency to tweet shortly before and after important events (such as TV discussions) which confirms previous findings from other elections \cite{Smailovic2015}. However, aside from that we also identified increased social media activity that did not correlate with the important events, but resulted from negative posts about a candidate. Moreover, we found that candidates tend to engage predominantly in a one-to-many communication pattern, and only occasionally respond directly to individual users. 

By applying natural language processing (NLP) techniques, we found evidence that both candidates used negative campaigning in order to get more supporters. In particular, our study distinguishes between misinformation and negative information, and we found that negative information in this particular presidential elections received a high re-tweet and like count. However, we also found evidence that the propagation of misinformation can backfire and have negative effects on the spreader. 

With respect to the tweeting behavior of the candidates' followers, we used sentiment analysis and network analysis techniques to construct ego-networks of the candidates, and examined which topics social media users associate with each candidate. Our results show a clear distinction in how users perceive both candidates. In particular, we found that hashtags carrying a negative connotation have predominantly been associated with Norbert Hofer. This phenomenon can be observed in both, the English and German language hashtag networks. Furthermore, in order to examine the communication patterns of social media users, we derived the corresponding communication networks. For example, we applied network analysis techniques to identify influential users in the German language network. In this context, we also showed that supporters of the two candidates tend to form two distinct clusters and predominantly communicate within their group (which might result in effects sometimes referred to as "filter bubble" or "echo chamber"). 

Similar to other technical infrastructures, social media platforms can be used for benign as well as malicious purposes. In particular, information spread over social networks has the potential to endanger democracy, disrupt markets, or manipulate people into joining radical groups \cite{WEF2013}. Thus, given the world-wide importance of social media in information seeking and opinion sharing, studying the way information is being disseminated over social media has become a field of significant economic, social, as well as technical impact. As part of our future work, we intend to study the impact of different emotions on the spread of information \cite{emotional-valence-shifts-asonam-2017, word-emotion-lexicons-snams-2017}. 

\newcommand{\etalchar}[1]{$^{#1}$}


\end{document}